\documentclass{article}
\newcommand{\ZZ}{\mbox{${\rm \:  Z\!\!\! I
\;\;}$}}

\def\vbar{\mathchoice{\vrule height6.3ptdepth-.5ptwidth.8pt\kern-.8pt}
  {\vrule height6.3ptdepth-.5ptwidth.8pt\kern-.8pt}
  {\vrule height4.1ptdepth-.35ptwidth.6pt\kern-.6pt}
  {\vrule height3.1ptdepth-.25ptwidth.5pt\kern-.5pt}}
\def\fudge{\mathchoice{}{}{\mkern.5mu}{\mkern.8mu}}
\def\bbc#1#2{{\rm \mkern#2mu\vbar\mkern-#2mu#1}}
\def\bbb#1{{\rm I\mkern-3.5mu #1}}
\def\bba#1#2{{\rm #1\mkern-#2mu\fudge #1}}
\def\bb#1{{\count4=`#1 \advance\count4by-64 \ifcase\count4\or\bba A{11.5}\or
  \bbb B\or\bbc C{5}\or\bbb D\or\bbb E\or\bbb F \or\bbc G{5}\or\bbb H\or
  \bbb I\or\bbc J{3}\or\bbb K\or\bbb L \or\bbb M\or\bbb N\or\bbc O{5} \or
  \bbb P\or\bbc Q{5}\or\bbb R\or\bbc S{4.2}\or\bba T{10.5}\or\bbc U{5}\or
  \bba V{12}\or\bba W{16.5}\or\bba X{11}\or\bba Y{11.7}\or\bba Z{7.5}\fi}}

\newcommand{\RR}{\mbox{${\rm \:  R\!\!\!\! I
\;\;}$}}

\newcommand{\vs}{\vspace{0.25cm}}

\newtheorem{theorem}{Theorem}
\newtheorem{itlemma}{Lemma}[section]
\newtheorem{itproposition}[itlemma]{Proposition}
\newtheorem{itcorollary}[itlemma]{Corollary}
\newtheorem{itremark}[itlemma]{Remark}
\newtheorem{itremarks}[itlemma]{Remarks}
\newtheorem{itdefinition}[itlemma]{Definition}
\newtheorem{itexample}[itlemma]{Example}

\newenvironment{lemma}{\begin{itlemma}\rm}{\end{itlemma}} 
\newenvironment{remark}{\begin{itremark}\rm}{\end{itremark}} 
\newenvironment{remarks}{\begin{itremarks} \rm}{\end{itremarks}}
\newenvironment{corollary}{\begin{itcorollary}\rm}{\end{itcorollary}}
\newenvironment{proposition}{\begin{itproposition}\rm}{\end{itproposition}}
\newenvironment{definition}{\begin{itdefinition}\rm}{\end{itdefinition}}
\newenvironment{example}{\begin{itexample}\rm}{\end{itexample}}
\newenvironment{fact}{\noindent {\em Fact}. \ \ }{\hfill \medskip}
\newenvironment{proof}{\noindent {\em Proof}.\ \ } {\hfill \medskip}
\newenvironment{claim}{\noindent {\em Claim}. \ \ }{\hfill \medskip}
\newcommand{\be}[1]{\begin{equation}\label{#1}}
\newcommand{\ee}{\end{equation}}
\newcommand{\bl}[1]{\begin{lemma}\label{#1}}
\newcommand{\br}[1]{\begin{remark}\label{#1}}
\newcommand{\brs}[1]{\begin{remarks}\label{#1}}
\newcommand{\bt}[1]{\begin{theorem}\label{#1}}
\newcommand{\bd}[1]{\begin{definition}\label{#1}}
\newcommand{\bp}[1]{\begin{proposition}\label{#1}}
\newcommand{\bc}[1]{\begin{corollary}\label{#1}}
\newcommand{\bfact}[1]{\begin{fact}\label{#1}}
\newcommand{\bex}[1]{\begin{example}\label{#1}}
\newcommand{\ec}{\end{corollary}}
\newcommand{\efact}{\end{fact}}
\newcommand{\eex}{\end{example}}
\newcommand{\el}{\end{lemma}}
\newcommand{\er}{\end{remark}}
\newcommand{\ers}{\end{remarks}}
\newcommand{\et}{\end{theorem}}
\newcommand{\ed}{\end{definition}}
\newcommand{\ep}{\end{proposition}}
\newcommand{\epr}{\end{proof}}
\newcommand{\bpr}{\begin{proof}}
\newcommand{\bcl}{\begin{claim}}
\newcommand{\ecl}{\end{claim}}

\newcommand{\bi}{\begin{itemize}}
\newcommand{\ei}{\end{itemize}}
\newcommand{\ben}{\begin{enumerate}}
\newcommand{\een}{\end{enumerate}}

\usepackage{graphicx}

\title{\bf \Large{Minimum Time Optimal Synthesis for a Control System on $SU(2)$}}

\vs

\vs

\author{Francesca Albertini\thanks{Dipartimento di Matematica,  Universit\`a di Padova, albertin@math.unipd.it}\, and \, Domenico D'Alessandro\thanks{Department of Mathematics, Iowa State University, Ames, Iowa, U.S.A., e-mail:daless@iastate.edu}}

\begin{document}

\maketitle

\vs
\vs
\begin{abstract}

For the  time optimal control on an invariant system on $SU(2)$, with two independent controls and a  bound on the norm of the control, the extremals of the maximum principle are explicit  functions of time. We use this fact here to perform the optimal synthesis for these systems, i.e., find all optimal trajectories.

Although the Lie group $SU(2)$ is three dimensional,  optimal trajectories can be described in the unit disk of the complex plane. We find that a circular trajectory separates optimal trajectories that reach the boundary of the unit disk from the others. Inside this separatrix circle another trajectory (the {\it critical trajectory}) plays an important role in that all optimal trajectories end at an intersection with this curve. 

Our results are of interest to find the minimum time needed to achieve a given evolution of a two level quantum system.

\end{abstract}

\section{Introduction}

The control of quantum mechanical systems has offered further motivation for the study of control systems on Lie groups, and in particular on $SU(n)$ and its Lie subgroups, as the evolution of a closed quantum system can be often modeled as a right invariant system varying on such Lie groups (see, e.g., \cite{IAAATACSI}, \cite{Mikobook} and references therein). Among these models, systems on $SU(2)$  arguably represent the simplest non trivial case, still a very rich one from a mathematical point of view. These {\it two-level quantum systems} are of fundamental interests in quantum physics and in quantum information, since they are the basic building block in the circuit based implementation of quantum information processing (see, e.g., \cite{NielsenandChuang}). A natural requirement in these implementations is to perform quantum operations (evolutions) in minimum time, both to shorten the overall time of computation and to avoid the effects of the interaction with the environment (de-coherence). For these reasons these systems have been studied in many aspects and their   (time) optimal control has been the subject  of many papers (see, e.g., \cite{Ugo1}, \cite{Ugo2},  \cite{newpaperonTimeOptimal}, \cite{newpaperbis}, \cite{QSL}, and references therein.). Here we add to this literature providing an explicit description of all optimal trajectories. This is done for a system with two orthogonal controls $u_x$ and $u_y$ (cf. model (\ref{basicmodel}) below) with have to satisfy a bound $u_x^2+u_y^2 \leq \gamma^2$ at every time, with positive $\gamma$ and $\gamma \leq 1$.

In particular, the model we consider is given by
\be{basicmodel}
\dot X=\sigma_z X + u_x \sigma_x X+u_y \sigma_y X, \qquad X(0)={\bf 1},
\ee
where $X \in SU(2)$ and $\sigma_{x,y,z}$ are the Pauli matrices, which form a basis of the Lie algebra $su(2)$. They are defined as
\be{Paulimat}
\sigma_x:=\frac{1}{2}\pmatrix{ 0 & i \cr i & 0}, \qquad
\sigma_y:=\frac{1}{2} \pmatrix{ 0 & -1 \cr 1 & 0},
\qquad \sigma_z:=\frac{1}{2}\pmatrix{ i & 0 \cr 0 & -i}.
\ee
The Lie algebra $su(2)$ is equipped with an inner product between  matrices, $\langle \cdot, \cdot \rangle$, defined as $\langle A, B \rangle:=Tr(A B^\dagger)$, so that the associated norm is $\|A\|:=\sqrt{\langle A, A \rangle}$. With these definitions the norm of the Pauli matrices is $\frac{1}{\sqrt{2}}$.

 We want to find, for every final condition $X_f$, the controls $u_x,u_y$, that steer the state of system (\ref{basicmodel}) from the identity to $X_f$ in minimum time, with the requirement that $u_x^2+u_y^2 \leq \gamma^2$, $\gamma^2 \leq 1$.

 \begin{remark}{\label{primo}}
Requiring a small bound on the norm of the control as compared to the size of the drift in (\ref{basicmodel}) is quite natural in NMR experiments where the control is usually a perturbation.
 \end{remark}

 \begin{remark}{\label{primobis}}
 The more general time optimal control problem for the system
\be{basicmodelgeneral}
\dot U=\pm \omega_0 \sigma_zU+v_x \sigma_x U+v_y \sigma_y U, \qquad U(0)={\bf 1},
\ee
with $\omega_0>0$ with $v_x^2+ v_y^2 \leq \omega_0^2 \gamma^2$ can be reduced to the problem for system (\ref{basicmodel}).  Define
$X(t):=U(\frac{t}{\omega})$, and new controls $u_{x,y}(t):=\frac{1}{\omega_0}v_{x,y}\left(\frac{t}{\omega_0}\right)$ we have that once the minimum time problem for
\be{piuomenomodel}
\dot X=\pm \sigma_zX+u_x \sigma_x X+u_y \sigma_y X, \qquad U(0)={\bf 1},
\ee
is solved with controls $u_x$ and $u_y$ and minimum time $T$, and $u_x^2+u_y^2 \leq \gamma^2$ the original optimal control for (\ref{basicmodelgeneral}) is solved with  $v_{x,y}(t)=\omega_0 u_{x,y}(\omega_0 t)$, in time $\frac{T}{\omega_0}$ to drive to the same final condition. The optimal control problem for system (\ref{piuomenomodel}) is the same as the one we have stated in the case $+$. In the case $-$ it can be reduced to it. Assume we have solved the minimum time problem for system (\ref{basicmodel}) for the final condition $X_f^{-1}$ and with controls $u_x$ and $u_y$ over an interval $[0,T]$. Then it is easily verified that the control $-u_x$, $-u_y$ over the same interval $[0,T]$ solves the problem of driving system (\ref{piuomenomodel}) with the $-$ from the identity to $X_f$, in minimum time.
\end{remark}
\vs

The paper is organized as follows. In section \ref{para}, we will select one of the methods to parametrize elements in $SU(2)$, and prove a simple property of the control system (\ref{basicmodel}) which will allow us to consider only two parameters rather than three when studying time optimal trajectories. In view of these facts, we will be able to perform the whole geometric analysis in the unit disk in the complex plane. We also recall how to apply the maximum principle of optimal control in this case and the form of the extremal controls and trajectories. In section \ref{Threecases} we solve the time optimal control problem for {\it diagonal operators}. As a limit of these trajectories, we identify a particular optimal trajectory which is a circle and plays a fundamental role for the whole analysis. All optimal trajectories leading to diagonal operators are outside this circle while all others are inside. Therefore we call this curve the {\it separatrix}.\footnote{Note this terminology is used with  a slightly different meaning usually in mathematics, where a separatrix is a curve separating different behaviors of solutions of a differential equation. Here our curves are projections of solutions of differential equations obtained for different values of parameters rather than initial conditions.}

For the special case $\gamma=1$, the separatrix curve coincides with the trajectory corresponding to the SWAP operator. The optimal trajectories for points outside the separatrix are the same ones that lead to diagonal operators. The optimal trajectories for points inside the separatrix are described  in section \ref{inside}. Here we give the general picture as a conjecture which is supported by theoretical results and simulations. In order to complete the proof though, we use the additional assumption $\gamma \geq \frac{1}{\sqrt{3}}$. In section \ref{Geo}, we provide a discussion of the results and show how these lead to a simple method to find the optimal control once the final condition is chosen. 
In this section we also compare our results with other work on the control of systems on $SU(2)$ and two level quantum systems and in particular \cite{newpaperonTimeOptimal} and \cite{newpaperbis}.

\section{Parametrization of $SU(2)$ and general properties of the model}\label{para}

\subsection{Parametrization of the final conditions in the optimal control problem}

It is well known that the Lie group $SU(2)$ is diffeomorphic to the sphere $S^3 \subseteq \RR^4$ and it is Lie-homeomorphic to the Lie group of unit quaternions, $SH$,  $x + y \vec i + c \vec j+ d \vec k$, with $x^2+y^2+c^2+d^2=1$, the homeomorphism being given by
\be{homeo}
SH \ni x + y \vec i + c \vec j+ d \vec k \Leftrightarrow \pmatrix{x+yi & -(c+id) \cr (c-id) & x-iy} \in SU(2).  \ee
By writing $-(c+id)=e^{i\phi}M$ and  $x+iy=e^{i\psi}\sqrt{1-M^2}$, with $0 \leq M \leq 1$, $\psi, \phi \in [0,2\pi)$, we can write any matrix $X_f\in SU(2)$ using the three parameters $\psi,\phi,$ and $M$, as
\be{parametriz}
X_f:=\pmatrix{e^{i\psi} \sqrt{1-M^2} & e^{i \phi} M \cr
- e^{-i\phi}M & e^{-i \psi} \sqrt{1-M^2}}.
\ee
We shall some times normalize the parameter $\psi$ and use the parameter $x_{\psi}$ instead, defined as $x_{\psi}:=\frac{\psi -\pi}{\pi}$, with  $x_\psi \in [-1,1)$.
The parameter $\phi$ of the final condition $X_f$ in (\ref{parametriz}) does not affect the time optimal control problem, in the sense that matrices that differ only by the parameter $\phi$ can be reached in the same minimum time. This is a consequence of the following proposition.

\bp{eliminaz} The minimum time to reach $X_f \in SU(2)$, is the same as the minimum time to reach
$e^{\sigma_z \alpha} X_f e^{-\sigma_z \alpha}$, for any $\alpha \in \RR$.
\ep
\bpr
Let $u_x$ and $u_y$ optimal controls steering the state $X$ of (\ref{basicmodel}) from the identity to $X_f$, in time $T_{opt}$ and let $X^o:=X^o(t)$ the corresponding trajectory. Define for $j=x,y$ the constants $\beta_{jk}$ such that
\be{rotation}
e^{\sigma_z \alpha}\sigma_{j}e^{-\sigma_z \alpha}=\sum_{k=x,y}\beta_{j,k} \sigma_k.
\ee
Define new controls $v_{x}, v_{y}$, for $k=x,y$, as  $v_{k}:=\sum_{j=x,y} \beta_{j,k} u_j$. Moreover notice that $v_x^2+v_y^2=u_x^2+u_y^2$ so that, if $u_x,u_y$ is an admissible control so is $v_x,v_y$. With the control $v_x,v_y$, the trajectory solution of (\ref{basicmodel}) is $U(t)=e^{\sigma_z \alpha} X^o(t) e^{-\sigma_z \alpha}.$ In fact, differentiating $U(t)$ and using (\ref{basicmodel}) for $X^o$ and (\ref{rotation}), we obtain
\[
\dot U:=e^{\sigma_z \alpha} \dot X e^{-\sigma_z \alpha}=\sigma_z U+(\sum_{j=x,y} u_{j}(\sum_{k=x,y} \beta_{j,k} \sigma_k)) U=
\]
\[\sigma_z U+(\sum_{k=x,y}(\sum_{j=x,y} \beta_{j,k}u_j) \sigma_k)U=
\sigma_z U+(\sum_{k=x,y}v_k \sigma_k)U.
\]
This shows that the optimal time to reach $e^{\sigma_z \alpha} X_f e^{-\sigma_z \alpha}$ is not greater than the one to reach $X_f$. By exchanging the roles of $X_f$ and $e^{\sigma_z \alpha} X_f e^{-\sigma_z \alpha}$, the opposite is seen to be true. Therefore the minimum time is the same in the two cases as stated.
\epr

\br{generalization}
The proof can be generalized with only formal modifications to more general systems on (Lie subgroups of) $SU(n)$, and more general systems of the form $\dot X=AX+\sum_{j=1}^m u_j B_j X$.  We can replace  the element of the form $e^{\sigma_z \alpha}$ with any element $K$  of (the Lie subgroup of) $SU(n)$, which commutes with $A$ and it such that $\texttt{span}\{ K B_1 K^\dagger,\ldots,  K B_m K^\dagger \}=\texttt{span}\{ B_1,\ldots, B_m\}$.
\er
\vs
In view of Proposition \ref{eliminaz} the only element that is relevant to determine the minimum time to reach $X_f$ in (\ref{parametriz}) is the element (1,1) in the matrix $X_f$. This will be parametrized by phase $\psi$ (or  $x_\psi$) and magnitude $M$ or, more often, by its real and imaginary parts, i.e., as a point $x+iy$ in the unit disk in the complex plane. To every (optimal)  trajectory in $SU(2)$ there corresponds a curve starting from (1,0) in the unit disk. Points in the unit disk correspond to classes of matrices in $SU(2)$ which can reached in the same minimum time.

\vs

\subsection{The Pontryagin maximum principle and the expression of optimal candidates}

\vs


 Consider the problem of driving the state $X$ of (\ref{basicmodel}) from the identity to a final condition $X_f$, with bound $u_x^2+u_y^2 \leq \gamma^2$, in minimum time.  The {\it Pontryagin Maximum Principle} states that, if $u_x,u_y$ is optimal, and $X_o$ is the optimal trajectory, then there exists a nonzero matrix $\tilde M \in  {su(2)}$, such that,  for almost every $t$, $u_x(t),$ $u_y(t)$, are the values of $v_x$ and $v_y$, that maximize the Hamiltonian function
 \be{PMPHam}
 H(\tilde M,X_o,v_x,v_y):=\langle \tilde M, X_o^\dagger \sigma_z X_o\rangle+v_x \langle \tilde M, X_o^\dagger \sigma_x X_o\rangle+v_y \langle \tilde M, X_o^\dagger \sigma_y X_o\rangle.
 \ee
Furthermore $H(\tilde M,X_o(t),u_x(t),u_y(t))$ is constant for almost
every $t$.

Define,
$b_{x,y,z}:= \langle \tilde M, X_o^\dagger \sigma_{x,y,z} X_o\rangle$. The maximization condition, implies that
\be{afterminimiz}
u_{x,y}=\gamma \frac{b_{x,y}}{\sqrt{b_x^2+ b_y^2}}
 \ee unless $b_x$ and $b_y$ are both zero, in which case the corresponding arc is  called {\it singular}.
Differentiating $b_{x,y,z}$ with respect to time, using (\ref{basicmodel}), and the standard commutation relations for the Pauli matrices\footnote{$[\sigma_x, \sigma_y]=\sigma_z,$ $[\sigma_y, \sigma_z]=\sigma_x,$ $[\sigma_z, \sigma_x]=\sigma_y.$}, we arrive at the following system of differential equations for $b_x,$ $b_y$ and $b_z$.
\be{bx}
\dot b_x=b_z u_y - b_y,
\ee
\be{by}
\dot b_y= b_x -b_z u_x,
\ee
\be{bz}
\dot b_z=b_y u_x-b_x u_y.
\ee
On a non singular arc, given the expression of the controls $u_x$ and
$u_y$ in (\ref{afterminimiz}) we have that $b_z$ is constant. This together with the fact that the Hamiltonian (\ref{PMPHam}), which takes the form $H=b_z+\gamma \sqrt{b_x^2+b_y^2}$, is also constant, implies that the controls $u_x$ and $u_y$ (for nonsingular extremals) can be written as (cf., the solutions of (\ref{bx}), (\ref{by}))
\be{controlli}
u_x=\gamma \sin(\omega t + \tilde \phi), \qquad u_y=-\gamma \cos(\omega t + \tilde \phi),
\ee
for some frequency $\omega \in \RR$ and phase $\tilde \phi \in \RR$. For singular arcs where $b_x \equiv b_y \equiv 0$, from (\ref{bz}) $b_z=const \not= 0$ which\footnote{If it was equal to zero it would imply $\tilde M=0$ which is excluded from the maximum principle.} therefore gives from (\ref{bx}), (\ref{by}), $u_x\equiv 0$, $u_y \equiv 0$. Therefore singular arcs starting from a point $X_1$ have the form $e^{\sigma_z t} X_1$, for $t\in [0,t_1]$ for some $t_1 >0$. We shall see in Theorem \ref{Diagsumma} and its proof that these arcs are never optimal.\footnote{General conditions to discard singular arcs are discussed in \cite{UgoKP} and the references therein.} Therefore in the optimal control problem we can restrict ourselves to nonsingular arcs.

\vs

Using the controls (\ref{controlli}) in (\ref{basicmodel}),  the resulting differential equation {\it can be explicitly integrated} (see, e.g., \cite{CT} p.446). Direct verification shows that the solution is given by
\be{soluzexpli}
X(t,\omega,\tilde \phi):=\pmatrix{ e^{i \omega \tau}(\cos(a \tau)+ i \frac{b}{a} \sin(a \tau)) & e^{i (\omega \tau + \tilde \phi)} \frac{\gamma}{a} \sin(a \tau) \cr  - e^{-i(\omega \tau + \tilde \phi)} \frac{\gamma}{a} \sin(a \tau)  & e^{-i \omega \tau}(\cos(a \tau)-i \frac{b}{a} \sin(a \tau))},
\ee
for $\tau=\frac{t}{2}$, $b:=1-\omega$, $a:=\sqrt{\gamma^2 + b^2}$.  For given $\omega$ and  $\tilde \phi$, the time  $T$ is the minimum time to reach if $X_f:=X(T,\omega,\tilde \phi)$ if there is no smaller  $T_1$ and pair $\omega_1$ and $\tilde \phi_1$ such that  $X_f:=X(T_1,\omega_1,\tilde \phi_1)$ .

\vs


In the expression (\ref{soluzexpli}), the phase of the element $(1,2)$ does not affect the (minimum) time to reach a given target, in the sense that we can always tune $\tilde \phi$ to give an arbitrary phase to the (1,2) element of the final condition, which provides an alternative way to prove Proposition \ref{eliminaz}.

\vs

{\bf Notation:} In the following we shall replace the notation $\tau$ with $t$, with the understanding that the new $`t'$ is half the $`t'$ we have mentioned so far.

\subsection{Properties of extremal curves}\label{prope}

Any candidate optimal is represented by a parametric curve in the complex plane, and in particular inside the unit disk, which starts from the point $(1,0)$ and represents the $(1,1)$ element of the trajectory of (\ref{basicmodel}). These curves can be parametrized by the frequency $\omega$ of the optimal control candidates while the phase does not play any role. They are explicitly given by (cf. (\ref{soluzexpli})

\be{curvex}
x(t):=x_{\omega}(t)=\cos(\omega t)\cos(at)-\frac{b}{a}\sin(\omega t)\sin(at),
\ee

\be{curvey}
y(t):=y_\omega(t)= \sin(\omega t)\cos(at)+\frac{b}{a}\cos(\omega t)\sin(at),
\ee

with $b:=1-\omega,$ $a=\sqrt{b^2+\gamma^2}.$

We also have (cf. (\ref{soluzexpli})) for the distance of the point from the origin,
\be{rquadro}
1-M^2(t):=r^2(t):=x^2(t)+y^2(t)=1-\frac{\gamma^2}{a^2}\sin^2(a t).
\ee
The phase $\psi(t)$ is given (cf. (\ref{soluzexpli})) for $0 \leq t \leq \frac{\pi}{2a}$ by
\be{fase1}
\psi(t)=\omega t + \arctan\left( \frac{b}{a} \tan(a t) \right),
\ee
and for $\frac{\pi}{2a} < t \leq  \frac{\pi}{a}$,
\be{fase2}
\psi(t)=\omega t+ \pi + \arctan\left( \frac{b}{a} \tan(a t) \right).
\ee

\vs

In the following there will be some values of the frequency $\omega$ which play an important role. We define them at the outset. In particular we define $\omega^*:\frac{1+\gamma^2}{2}$, $\omega_c:=2 \omega^*=1+\gamma^2$. Correspondingly, we define $b^*:=1-\omega^*$, $b_c:=1-\omega_c$,
$a^*:\sqrt{\gamma^2+(b^*)^2}$ and $a_c:=\sqrt{\gamma^2+(b_c)^2}$.

We record few  properties of the extremal trajectories.

\vs

{\bf Fact 1} From equation (\ref{rquadro}), we have:
\[
\frac{d r^2}{dt}= \frac{-2\sin(at)\cos(at)}{a},
\]
which implies, that $r(t)$ is decreasing for $t\in[0,\frac{\pi}{2a}]$, and it is increasing for $t\in [\frac{\pi}{2a},\frac{pi}{a}]$. At the time $t=\frac{\pi}{a}$ the trajectory reaches the boundary of the unit disk.
Moreover, since $\frac{d(r^2)}{dt}|_{t=0}=\frac{d^2(r^2)}{dt^2}|_{t=0}=\frac{d^3(r^2)}{dt^3}|_{t=0}=0$ and $\frac{d^4(r^2)}{dt^4}|_{t=0} =8a^2$,  we have that given $\omega_1$ and $\omega_2$
and letting $a_{1,2}$  and  $r_{1,2}(t)$ the corresponding value for the constant $a$ and $r(t)$,
if $a_1>a_2$,  for $t$ in a  neighborhood  of $0$, we have $r_1(t)>r_2(t)$.

\vs

{\bf Fact 2}  Calculating $\frac{d \psi}{d t}$ from (\ref{fase1}), (\ref{fase2}), we obtain
\be{dpsidtau}
\frac{d \psi}{d t}=\frac{-\gamma^2 \sin^2(a t)\omega + a^2}{a^2 \cos^2( a t)+b^2 \sin^2(a t)}.
\ee
Equation (\ref{dpsidtau}) implies that  for $\omega \leq 1$ the phase is always increasing.
Moreover when $\omega>0$, we have:
\[
-\gamma^2 \sin^2(a \tau)\omega + a^2= \omega^2-(2+\gamma^2\sin^2(at))\omega+ (\gamma^2+1)
\geq \omega^2-(2-\gamma^2)\omega+(\gamma^2+1).
\]
Since the last polynomial is positive when $\omega>\omega_c:=1+\gamma^2$, we derive that the phase is increasing
for $\omega\leq 1$ and for $\omega\geq\omega_c$.

\vs

{\bf Fact 3}: Because of the {\it existence}  of the optimal control, every point in the unit disk is reached by at least one  curve and among those that reach the point at least one  is optimal.

\vs

{\bf Fact 4}: The {\it singular} curve corresponds to the {\it boundary} of the unit disk. Therefore  every point in the interior of the unit disk must be reached by an optimal trajectory  which contains a nonsingular arc. We shall in fact see in Theorem \ref{Diagsumma} (under the assumption $\gamma \leq 1$) that even for the points on the boundary the optimal trajectories are nonsingular, and this implies that all the optimal trajectories do not contains singular arcs.

\vs

{\bf Fact 5}: (Principle of Optimality) If a curve reaching a point $P$ is optimal, then that curve is optimal for every point on that curve before $P$.

\vs

{\bf Fact 6}: When two curves intersect at a point $P$ they cannot be both optimal at the point $P$. In fact, if they reach $P$ at point at different times, then, obviously, the one that reaches at greater time is not optimal. If the reach $P$ at the same time, then we could possibly switch from one value of $\omega$ to the other in the control and still have an optimal trajectory. This contradicts the fact that all the nonsingular extremals have the form (\ref{controlli}) (cfr. Fact 4).

If a curve is optimal for every point before a point $P$ and not optimal after $P$ we say that a curve {\it looses optimality at $P$}.

\section{Optimal control problem for diagonal final conditions and the separatrix curve}\label{Threecases}
%

\subsection{Diagonal operators}\label{operatoridiagonali}

Assume the final condition $X_f:=\pmatrix{e^{i\psi_f} & 0 \cr 0 & e^{-i \psi_f}}$, that is, we want to drive in minimum time to a point on the boundary of the unit disk. According to formula (\ref{rquadro}) extremal trajectories reach the boundary of the unit disk at times $T=\frac{k\pi}{a}$.  If $T$ is the final time in (\ref{soluzexpli}), we have the two equations
\be{equa1}
T=\frac{k \pi}{a}, \qquad k \geq 0,
\ee
\be{equa2}
\omega T+aT=\psi_f +2m\pi, \qquad m \in \ZZ,
\ee
which give respectively the condition on the norm of the off diagonal term and on the phase of the diagonal term.\footnote{In the condition (\ref{equa1}) we have used the fact that the time has to be nonnegative (in fact positive if $\psi \not=0$).} Plugging (\ref{equa1}) into (\ref{equa2}), we have
\be{equa3}
k \pi (1+ \frac{\omega}{a})=\psi_f + 2 m \pi.
\ee
A study of the function $f(\omega):=\frac{\omega}{a}$ for $\omega \in (- \infty, \infty)$ reveals that this function is bounded below by $-1$, so that, when  $\psi_f \in (0,2 \pi)$,  (\ref{equa3}) can only be verified for $m \geq 0$. Let use denote by $T_{k,m}$ the time $T$ which is given by equation (\ref{equa1}) with $k$ and verifying the constraint (\ref{equa2}). Notice that not all pairs $k>0,m\geq 0$ are feasible (the function $\frac{\omega}{a}$ is bounded).  We shall show that no matter what $\psi_f \in (0,2\pi)$ is,  the minimum of these  times is $T_{1,0}$ which is feasible.\footnote{This means that there exists an $\omega$ satisfying (\ref{equa3}) with $k=1$ and $m=0$.}.
%
The proof  can be achieved in two steps given by the following two lemmas. The result for the diagonal case is summarized in Theorem \ref{Diagsumma}. Proofs of Lemmas \ref{Lemma1} and \ref{Lemma2} are given in Appendix A.

\bl{Lemma1} For every $k>0$ and $m>0$,
\be{TKMTK0}
T_{k,m} \geq T_{k,0}.
\ee
\el

\bl{Lemma2} For every $k >0$,
\be{TK0T10}
T_{k,0} \geq T_{1,0}.
\ee
\el

\bt{Diagsumma}
Assume $\gamma \leq 1$. Then the minimum time to reach a diagonal operator $X_f :=\pmatrix{e^{i \psi_f} & 0 \cr
0 & e^{-i \psi_f}}$, $\psi_f \in (0, 2 \pi)$  is
\be{Tmin}
T_{min}=T_{1,0}(\psi_f):=\frac{\psi_f (2 \pi - \psi_f)}{\pi - \psi_f +\sqrt{\pi^2 + \gamma^2 \psi_f (2\pi - \psi_f)}},
\ee
which is obtained with the controls  (\ref{controlli}), with $\tilde \phi$ arbitrary and $\omega$ given by\footnote{Recall that $x_\psi:=\frac{\psi -\pi}{\pi}$}
\be{omegaopt}
\omega= \frac{x_{\psi_f}}{1- x^2_{\psi_f}} (- x_{\psi_f}+ \sqrt{1+ \gamma^2(1-x^2_{\psi_f})}).
\ee
\et
\bpr
The theorem summarizes the previous two Lemmas. The expression of the optimal
frequency $\omega$ is obtained from (\ref{equa3}), (\ref{aa}), with $k=1$ and $m=0$.

To make sure that this time is optimal we need to compare it with the one obtained with the singular trajectory which is $T_{sing}(\psi_f)=\psi_f$. In fact we have $T_{min}< T_{sing}$. This follows  from
\be{singula00}
\frac{T_{min}}{\psi_f} = \frac{2\pi - \psi_f}{\pi - \psi_f + \sqrt{\pi^2 + \gamma^2 \psi_f (2 \pi - \psi_f)}}< 1=\frac{T_{sing}}{\psi_f}.
\ee
\epr

A consequence of this theorem is also that no optimal trajectory  can be contain a singular arc, because the singular arc can be followed in smaller time. 

\vs

\subsection{The separatrix curve}

Reconsider formula (\ref{omegaopt}). There is a one to one correspondence between values of $x_{\psi_f} \in (-1,1)$ (alternatively values of $\psi_f \in (0,2\pi)$) and $\omega$. In fact $-\infty <\omega < \frac{1+\gamma^2}{2}:=\omega^*$ and $\lim_{\psi_f \rightarrow 0} \omega =-\infty$ and  $\lim_{\psi_f \rightarrow 2 \pi} \omega =\frac{1+\gamma^2}{2}=\omega^*$.

Consider now the trajectory corresponding exactly to $\omega= \omega^*=\frac{1+\gamma^2}{2}$. In this case
$a=a^*=\omega=\omega^*$ and the parametric equations  of (\ref{curvex}) and (\ref{curvey}) become
\be{curvexsepa}
x(t)=\frac{2}{1+\gamma^2} \cos(\omega^*t)-\frac{1-\gamma^2}{1+\gamma^2},
\ee
\be{curveysepa}
y(t)=\frac{2}{1+\gamma^2} \cos(\omega^*t)\sin(\omega^*t).
\ee
This represents a circle with center in
\be{centro}
P=\left(\frac{\gamma^2}{1+\gamma^2}, 0 \right),
\ee
and radius $\frac{1}{1+\gamma^2}$. We shall call this circle the `{\it separatrix}'. The following lemma justifies this  name.
\bl{stannofuori}
All the optimal trajectories corresponding to diagonal operators (described in subsection \ref{operatoridiagonali}) intersect the separatrix curve only in the point $(1,0)$.
\el

The proof is in Appendix A.

Figures \ref{Traie12} and \ref{Traie1} give some plots of the trajectories outside the separatrix, leading to diagonal operators for the cases $\gamma=\frac{1}{2}$ and $\gamma=1$ respectively. The separatrix is the red circle in both cases. The cases $\omega=-3$, $\omega=0$ and $\omega=\frac{1}{2}$ are displayed explicitly for $\gamma=\frac{1}{2}$ and the same values of $\omega$'s and $\omega=\frac{8}{9}$ are displayed for $\gamma=1$. As $\omega \rightarrow \omega^*$, the trajectories tend to the separatrix.

\begin{figure}[htb]
\centering
\includegraphics[width=0.7\textwidth]{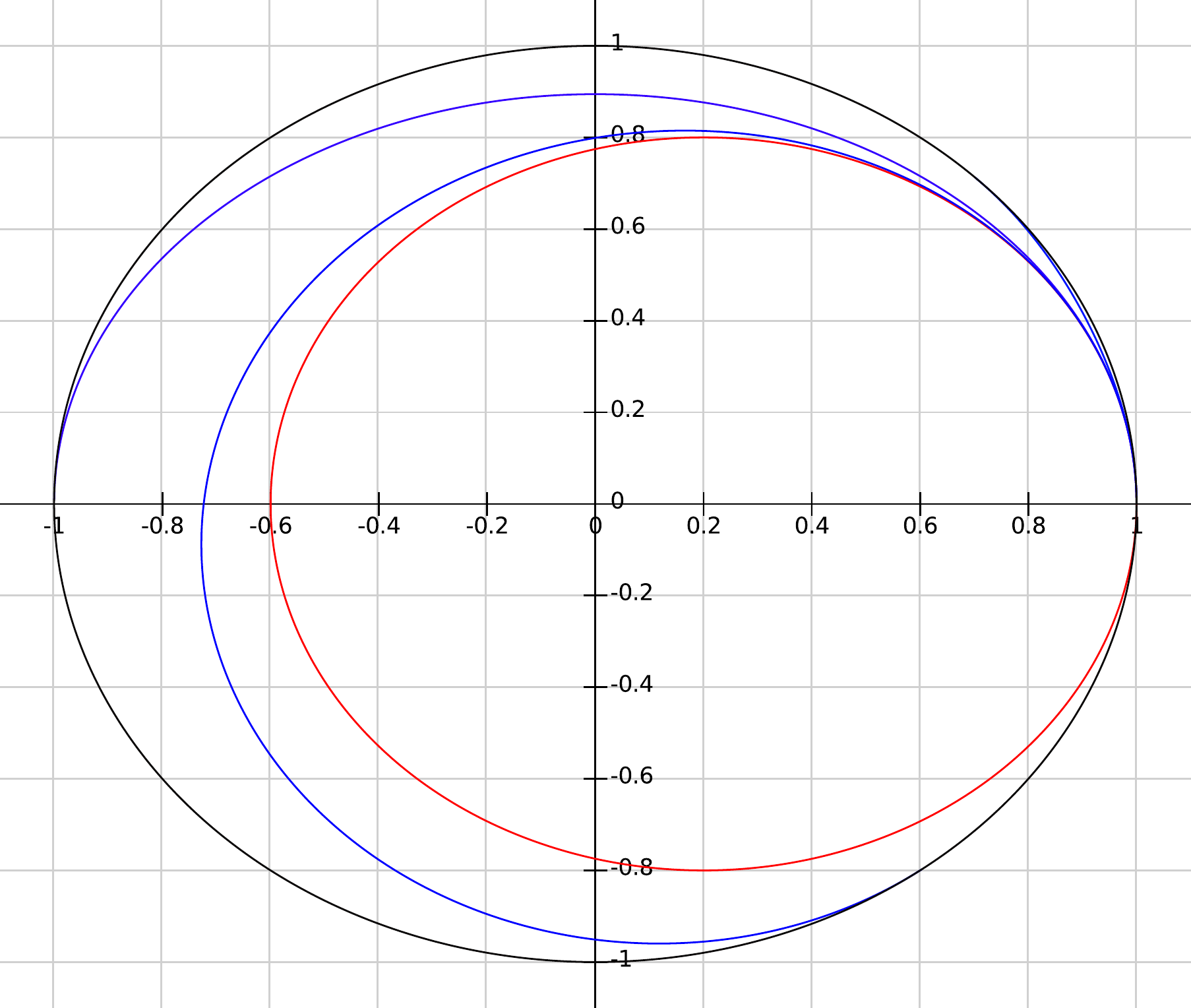}
\caption{Optimal trajectories to reach the boundary of the unit disk  (in blue) for various values of $\omega$ and $\gamma=\frac{1}{2}$. The outermost trajectory is the one corresponding to $\omega=-3$, the next one (reaching the point (-1,0)) corresponds to $\omega=0$. The innermost trajectory is the one corresponding to $\omega=\frac{1}{2}$. The separatrix is the red circle centered at the point $(\frac{1}{5}, 0)$.}
\label{Traie12}
\end{figure}

\begin{figure}[htb]
\centering
\includegraphics[width=0.7\textwidth]{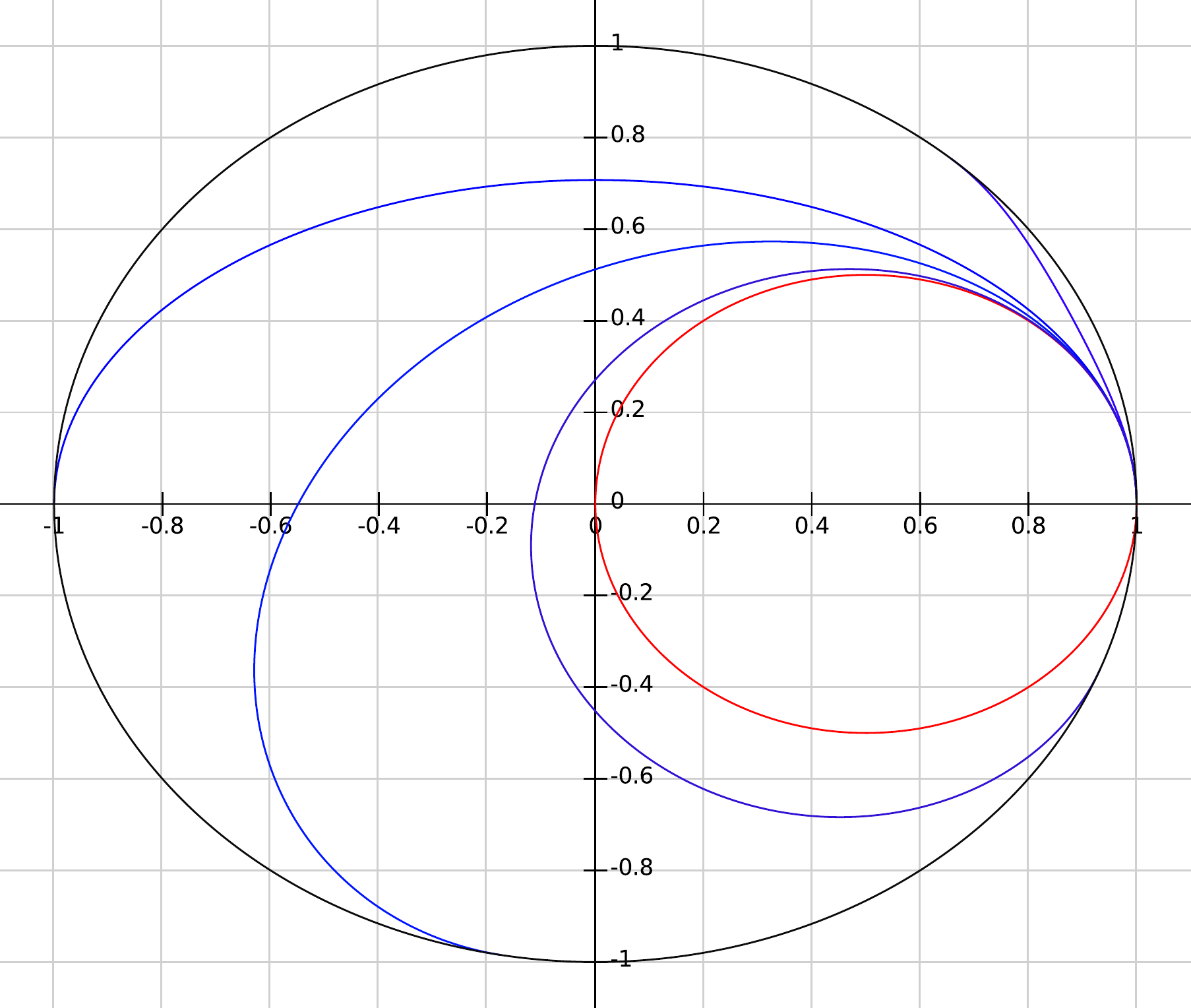}
\caption{Optimal trajectories to reach the boundary of the unit disk  (in blue) for various values of $\omega$ and $\gamma=1$. The outermost trajectory is the one corresponding to $\omega=-3$, the next one (reaching the point (-1,0)) corresponds to $\omega=0$. The next on is the one corresponding to $\omega=\frac{1}{2}$. The innermost trajectory is the one corresponding to $\omega=\frac{8}{9}$. The separatrix is the red circle centered at the point $(\frac{1}{2}, 0)$.}
\label{Traie1}
\end{figure}

The following proposition states two important properties of the optimal trajectories outside the separatrix.

\bp{Filling}

\begin{enumerate}

\item The trajectories corresponding to $\omega \in (-\infty, \omega^*)$ loose optimality after reaching the boundary of the unit disk.

\item Every point outside the separatrix is reached by an optimal trajectory (before reaching the boundary) corresponding to a single  value of $\omega$, with $\omega \in (-\infty, \omega^*)$

\end{enumerate}

\ep

\bpr
To prove 1., recall from  Fact 2 of subsection \ref{prope}  that the phase $\psi$ is always increasing, since $\omega^*\leq 1$. This means that any of the  trajectories corresponding to $\omega \in (-\infty, \omega^*)$  after hitting the boundary will necessarily intersect another trajectory corresponding to a larger value of (final) $\psi_f$ which is optimal. Therefore such a trajectory looses optimality at the boundary.

To prove 2.,  Consider a point $P$ outside the separatrix and assume by contradiction that none of the curves reaching the boundary and corresponding to  $\omega \in (-\infty, \omega^*)$ contains such a point. In particular, denote by $C_{\psi_f}$ any such curve corresponding to the phase $\psi_f \in (0, 2\pi)$. By the existence of the optimal control for $P$ there exists an optimal trajectory ending in  $P$, which we denote by $C_P$, defined in $[0,t_P]$, with $t_P <\frac{\pi}{a}$ (see (\ref{rquadro}). All the trajectories $C_{\psi_f}$ and $C_P$ never intersect (except for the point $(1,0)$).
Express the trajectory $C_{\psi_f}$ and $C_P$ as polar equations $r=r(\psi)$ with $\psi$ the (variable) phase. In particular we write $r=r_f( \psi)$ for $C_{\psi_f}$ and $r=r_P( \psi)$ for $C_P$. With this notation, we say that $C_{\psi_f}$ is {\it above} $C_P$ if $r_f( \psi)$ is greater than $r_P(\psi)$ for one (and therefore all since they cannot intersect) $\psi \not=0$ which are in the common domain of the function $r_f$ and $r_P$. Analogously we say that  $C_{\psi_f}$ is {\it below} $C_P$ if $r_f(\psi)$ is smaller than $r_P(\psi)$. Consider the set $A_P$ ($B_P$) of all $\psi_f \in (0,2\pi)$ which are such that $C_{\psi_f}$ is above (below) $C_P$. It is important to notice that both $A_P$ and $B_P$ are not empty. $A_P$ is not empty because it definitely contains all $\psi_f$'s smaller than the phase of $P$ since the phase is always increasing from  formula (\ref{dpsidtau}). $B_P$ is not empty because it is enough to take a curve $C_{\psi_f}$ sufficiently close to the separatrix to leave $P$ on the right.    Moreover $A_P \bigcup B_P =(0,2\pi)$. By continuity (again using the fact that $C_{\psi_f}$ and $C_P$ never intersect) $A_P$ and $B_P$  are both  open set. Since they are not empty this contradicts $A_P \bigcup B_P =(0,2\pi)$ because of the connectedness of $(0,2\pi)$. \epr

\subsubsection{SWAP operator}{\label{swap}}

\begin{figure}[htb]
\centering
\includegraphics[width=0.7\textwidth]{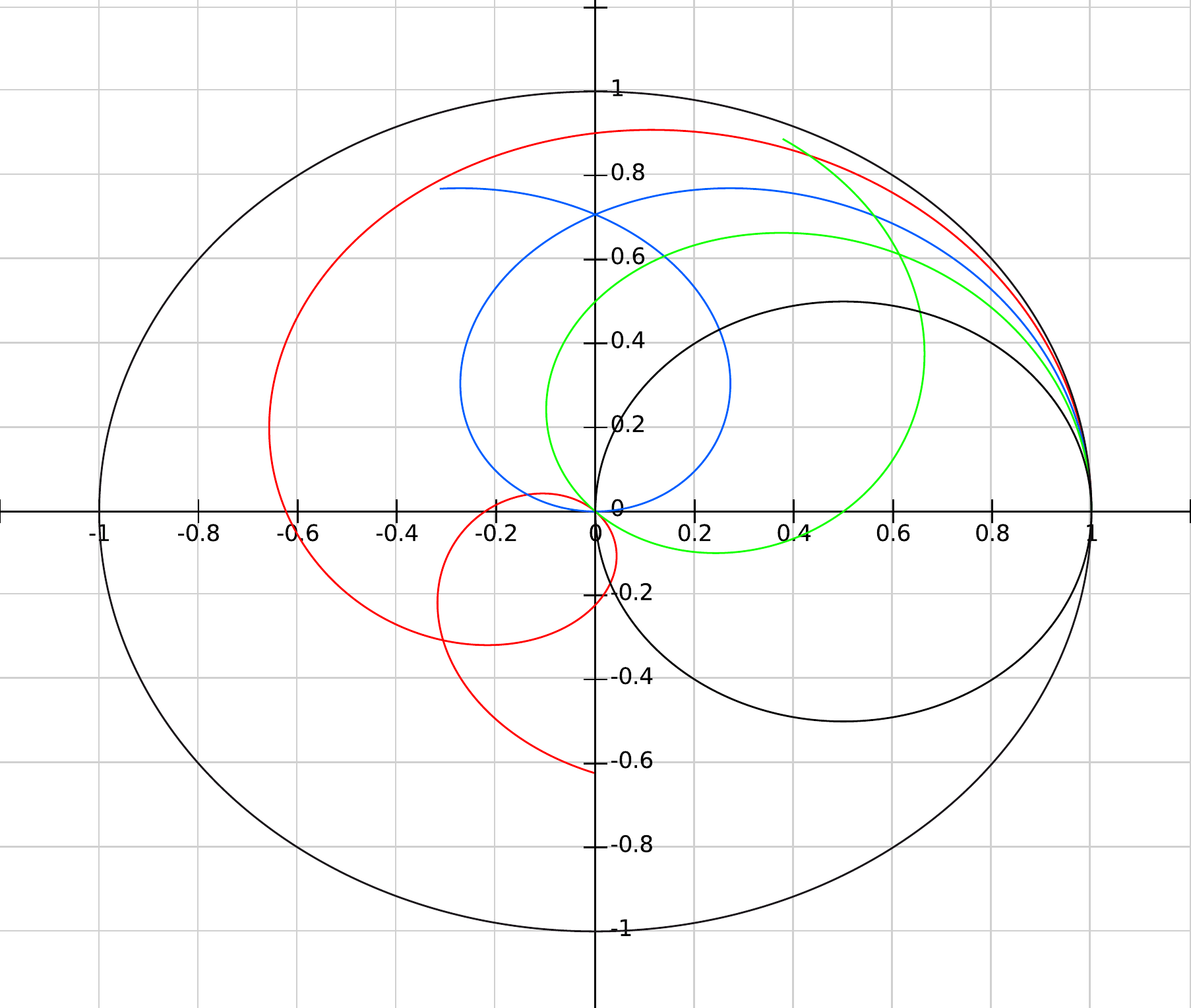}
\caption{Optimal trajectories for the SWAP operator for various values of $\gamma$. Trajectories are no longer optimal after they self-intersect. The trajectories correspond to the following values of $\gamma$: $\gamma=\frac{2}{7}$, red; $\gamma=\frac{1}{2}$, blue; $\gamma=\frac{2}{3}$, green; $\gamma=1$, black.}
\label{figurewithvariousgamma}
\end{figure}

The SWAP operator, is the operator that in quantum information theory corresponds to a logic operation $NOT$. It inverts the state of a two level quantum system. It is given in the computational basis by
\be{SWAP}
X_{SWAP}:=\pmatrix{0 & 1 \cr -1 & 0},
\ee
which corresponds to the origin of the unit disk.  In formula (\ref{soluzexpli}), we need $a=\gamma$, $b=0$ and $\omega=1$ (resonance condition \cite{UgoKP}) and, minimum time $T_{min}(X_{SWAP})=\frac{\pi}{2\gamma}$. The optimal trajectory is
\be{OptSWAP}
X(t)=\pmatrix{e^{i t} \cos(\gamma t) & \sin (\gamma t) \cr - \sin(\gamma t) & e^{-i t} \cos(\gamma t)}.
\ee
Figure \ref{figurewithvariousgamma} displays the various trajectories for the values of $\gamma=\frac{2}{7},\frac{1}{2},\frac{2}{3},1$, until the trajectories self intersect and therefore are no longer optimal.\footnote{The full  trajectories are closed in the case where $\gamma$ is a rational number.} The trajectory corresponding to $\gamma=1$ is a circle of radius $\frac{1}{2}$ centered at $(\frac{1}{2},0)$. Which coincides with the separatrix in this case.

\section{The optimal control problem inside the separatrix}\label{inside}

From now on we denote by $\mathcal{S}$ the closed region inside the separatrix. 
For points in $\mathcal{S}$, the frequency $\omega$ of the optimal control must be  greater than or equal to $\omega^*=\frac{1+ \gamma^2}{2}$. In fact, as we have seen in the provious section,
the trajectories corresponding to $\omega < \omega^*$ do not intersect the separatrix  before touching the boundary (Lemma \ref{stannofuori}) and, after touching the boundary, they are not optimal anymore (Proposition \ref{Filling}). Therefore, for all points in $\mathcal{S}$, the optimal trajectories are with omega, $\omega \geq \omega^*$.

In order to study the behavior of these trajectories with respect to the separatrix we consider a trajectory $(x_\omega(t), y_\omega(t))$ in (\ref{curvex}), (\ref{curvey}) and the function
\be{funzionedistanza}
\Delta_\omega(t):=\left(x_\omega(t)-\frac{\gamma^2}{\gamma^2+1}\right)^2+ y_\omega^2(t)-\frac{1}{(1+\gamma^2)^2},
\ee
which gives the difference between the square of the distance of the trajectory (as a function of $t$) from the center of the separatrix and the square of the radius of the separatrix. $\Delta_\omega(t)$ is identically zero for $\omega=\omega^*$, i.e., on the separatrix. Using (\ref{rquadro}) and (\ref{curvex}), we obtain
\be{iii}
\frac{\Delta_\omega(t)}{\gamma^2}=\frac{2}{1+\gamma^2}-\frac{1}{a^2}\sin^2(a t)-\frac{2}{1+\gamma^2}(\cos(\omega t) \cos(a t)- \frac{b}{a}\sin(\omega t) \sin(a t)).
\ee
\bl{lemmino}
Assume $\omega \in [\omega_*, 3 \omega^*)$. Then there exists an
$\epsilon=\epsilon_\omega > 0$ such that $(x_\omega(t), y_{\omega}(t))$ is in $\mathcal{S}$ for every $t\in [0, \epsilon)$. Assume $\omega > 3 \omega^*$. Then there exists an $\epsilon=\epsilon_\omega$ such that $(x_\omega(t), y_{\omega}(t))$ is outside  $\mathcal{S}$ for every $t\in (0, \epsilon)$.
\el
\bpr
We calculate the derivatives of $\frac{\Delta_\omega(t)}{\gamma^2}$ at $t=0$. The first three derivatives give zero while the fourth one is greater than or equal to zero for $\omega \leq \omega^*$ and $\omega \geq \omega^*$ otherwise it is smaller than zero. The case $\omega \leq  \omega^*$ corresponds to the  trajectories of subsection \ref{operatoridiagonali} and the separatrix itself. The case $\omega >  \omega^* $ also corresponds to trajectories that starts outside of the separatrix. Trajectories corresponding to $\omega \in (\omega_*, 3 \omega^*)$ start inside the separatrix.
\epr
\bc{exclusione}
Trajectories corresponding to $\omega > 3 \omega^*$ are not optimal.
\ec
\bpr
Using Proposition \ref{Filling}, these trajectories are not optimal since they intersect the optimal ones going to the boundary of the unit disk.
\epr

From the above two results, all points in $\mathcal{S}$ will have optimal trajectories corresponding  to values of $\omega$ in the interval $[\omega^*, 3 \omega^*]$.

\vs

In the interval $[\omega^*,3 \omega^*]$ a particularly important role is played by the curve corresponding to $\omega_c:=2 \omega^*=\gamma^2+1$. This curve presents a cuspid point, i.e., a point where both $\dot x$ and $\dot y$ are zero. In particular having defined  $a_c$ as the value of $a$ corresponding to $\omega_c$, i.e., $a_c:=\gamma \sqrt{1+\gamma^2}$, from (\ref{curvex}) and (\ref{curvey}), we obtain
\be{derivatx}
\dot x_{\omega_c}(t)=-\sin(\omega_c t) \cos(a_c t),
\ee
\be{derivaty}
\dot y_{\omega_c}(t)=\cos(\omega_c t) \cos(a_c t),
\ee
and both derivatives are zero when $t =\frac{\pi}{2 a_c}$. We shall call the trajectory corresponding to $\omega=\omega_c$ until the point corresponding to $t =\frac{\pi}{2 a_c}$, the {\it critical trajectory}. Its final point is
\be{finalcritic}
x_{\omega_c}(\frac{\pi}{2 a_c})=\frac{\gamma}{\sqrt{1+\gamma^2}}\sin\left(\pi \frac{\sqrt{1+\gamma^2}}{2\gamma}\right), \qquad y_{\omega_c}(\frac{\pi}{2 a_c})=- \frac{\gamma}{\sqrt{1+\gamma^2}}\cos\left(\pi \frac{\sqrt{1+\gamma^2}}{2\gamma}\right).
\ee
It is in particular a point on the circle centered at the origin with radius $\frac{\gamma}{\sqrt{1+\gamma^2}}$. Such a circle centered at the origin will play an important role in our proof below. We call it the {\it critical circle}.

\vs

The general picture of the optimal synthesis for points inside the separatrix is summarized  in Theorem \ref{Insideseparatrix}.


\bt{Insideseparatrix}

\begin{figure}[htb]
\centering
\includegraphics[width=0.7\textwidth]{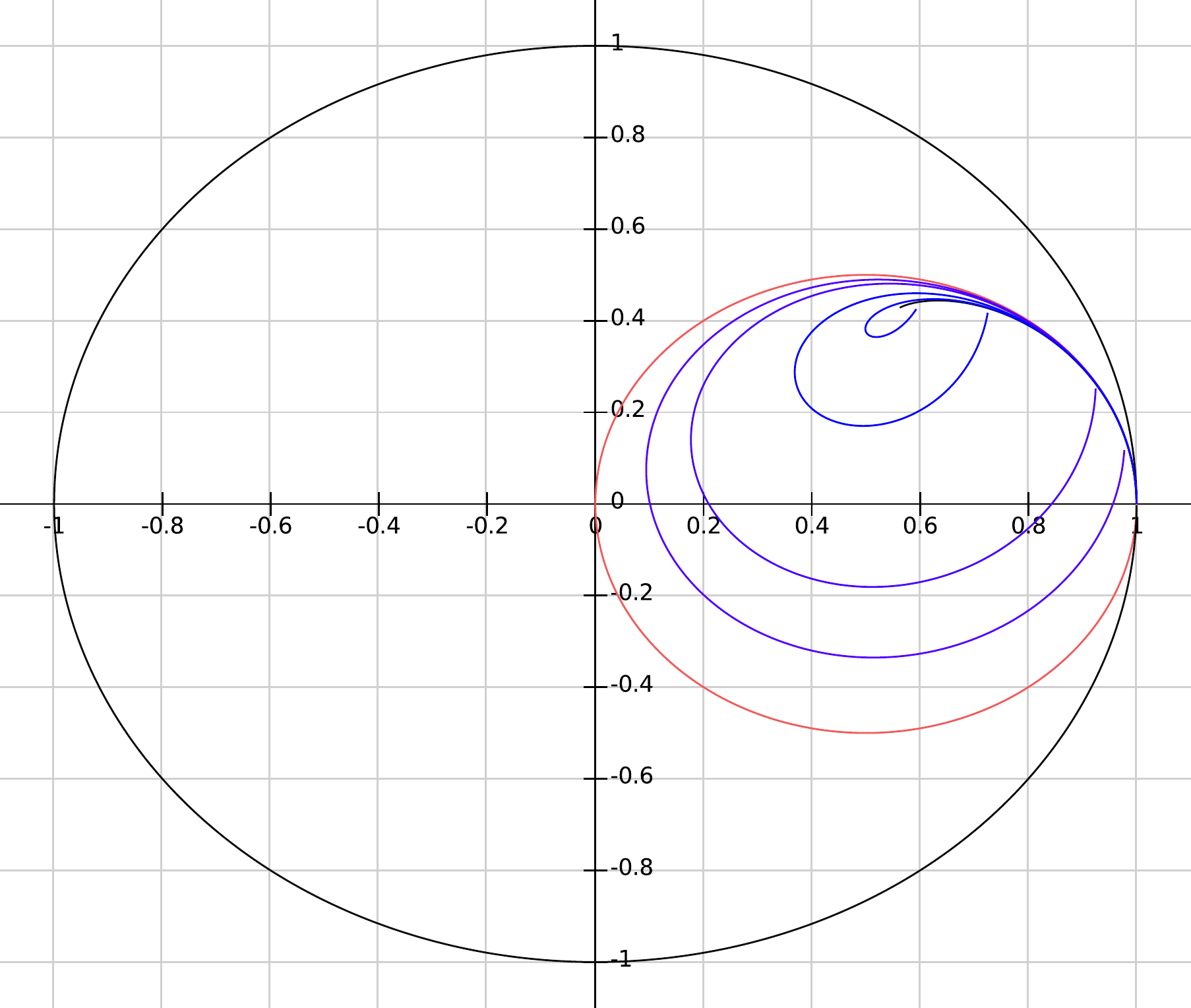}
\caption{Optimal trajectories inside the separatrix (in red) for $\gamma=1$. The critical
trajectory is in black, while the trajectories for $\omega=1.1 \omega^*$,
$\omega=1.2 \omega^*$, $\omega=1.5 \omega^*$, $\omega=1.8 \omega^*$, are in blue (starting closer to the separatrix when $\omega \rightarrow \omega^*=1$ and starting closer to the critical trajectory when $\omega \rightarrow \omega^*=2$).}
\label{Inssepa1}
\end{figure}

\begin{figure}[htb]
\centering
\includegraphics[width=0.7\textwidth]{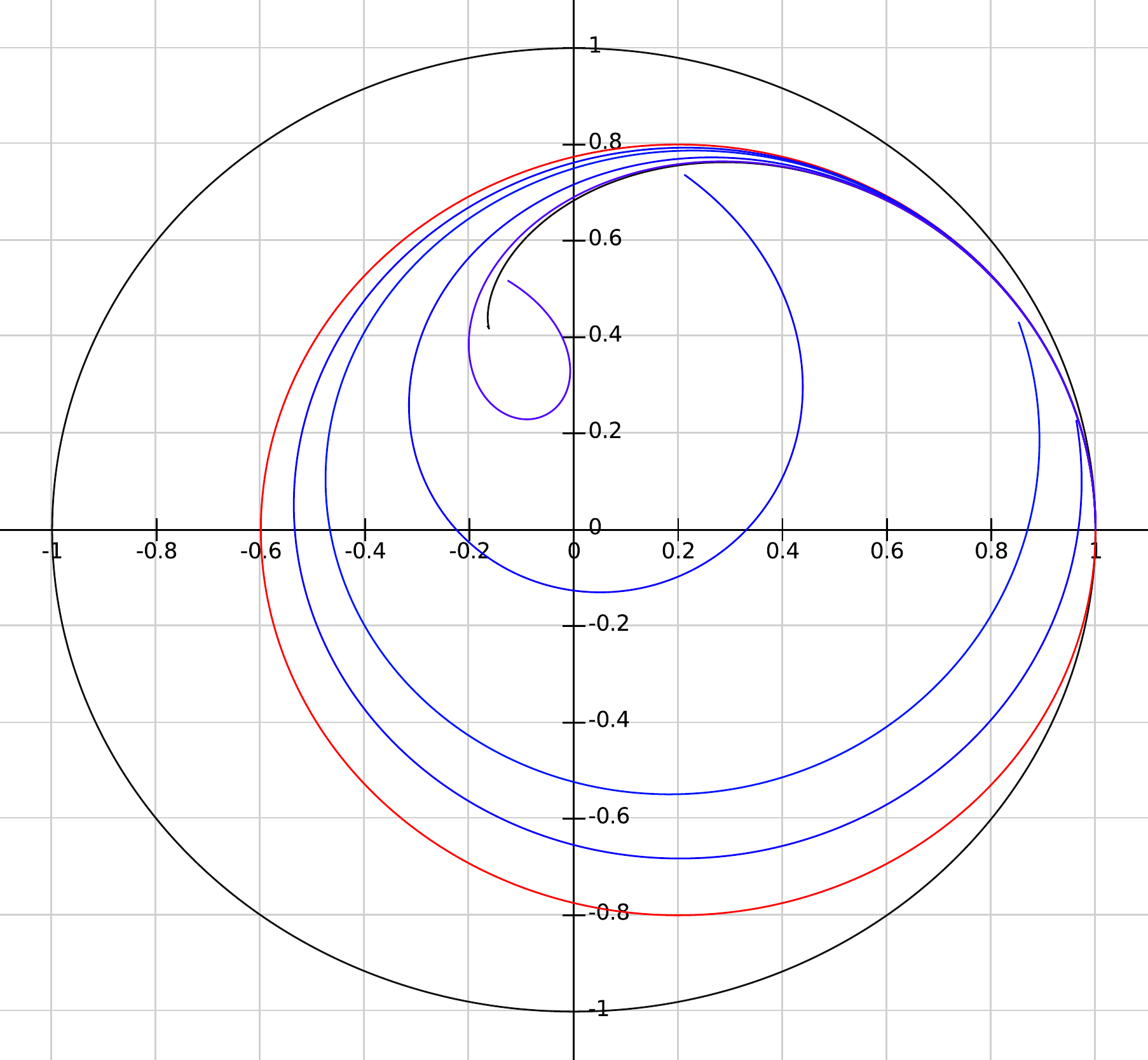}
\caption{Optimal trajectories inside the separatrix (in red) for $\gamma=\frac{1}{2}$. The critical
trajectory is in black, while the trajectories for $\omega=1.1 \omega^*$,
$\omega=1.2 \omega^*$, $\omega=1.5 \omega^*$, $\omega=1.8 \omega^*$, are in blue (starting closer to the separatrix when $\omega \rightarrow \omega^*=\frac{1+\gamma^2}{2}=
\frac{1+\frac{1}{4}}{2}$ and starting closer to the critical trajectory when $\omega \rightarrow 2\omega^*=1+\gamma^2=\frac{5}{4}$).}
\label{Inssepa2}
\end{figure}

\begin{figure}[htb]
\centering
\includegraphics[width=0.7\textwidth]{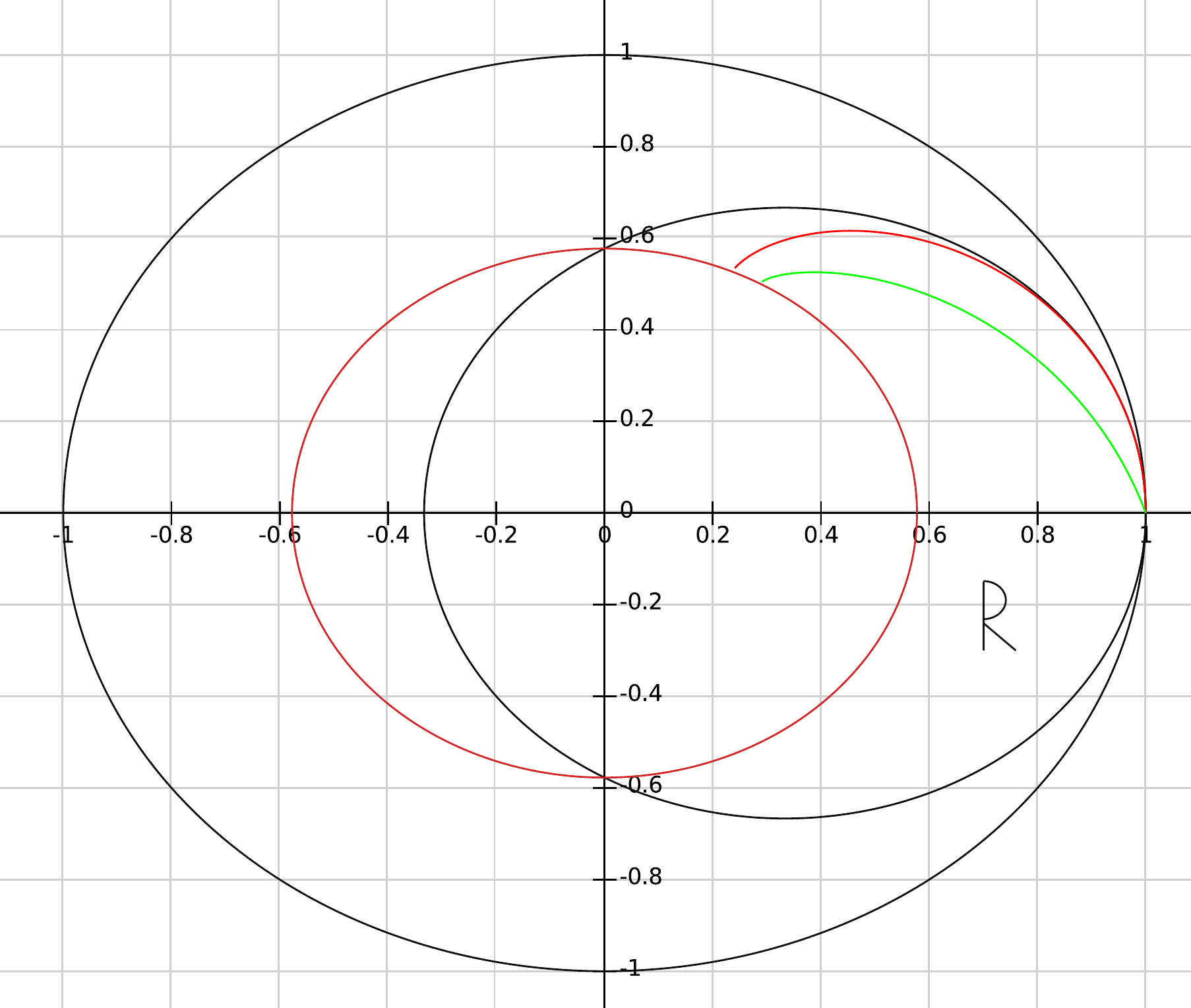}
\caption{Geometric objects used in the proof of Theorem \ref{Insideseparatrix}. The critical trajectory and the critical circle are in red ($\gamma=\frac{1}{sqrt{2}}$). We deform the critical trajectory by adding a $\-epsilon \lambda$ to the phase. The corresponding deformed curve is in green. Trajectories for $\omega\in (2 \omega^*, 3 \omega^*]$ never reach the region $R$ in the figure.}
\label{Perlaprova}
\end{figure}

Assume $\gamma \in [\frac{1}{\sqrt{3}}, 1]$. The only optimal trajectories for points in $\mathcal{S}$ correspond to $\omega \in [\omega^*, 2 \omega^*]$. The trajectory corresponding to $\omega^*$ is the optimal for points of the separatrix. The trajectory corresponding to $\omega_c=2 \omega^*$ until the point (\ref{finalcritic}) is optimal for point on the critical trajectory. For any other points inside the separatrix, there exists a unique value of $\omega \in (\omega^*, 2\omega^*)$ and a corresponding optimal trajectory leading to that point.
\et

We believe this theorem holds for general values of $\gamma \leq 1$ but we were able to completely  prove  it only  for  $\gamma \in [\frac{1}{\sqrt{3}}, 1]$. The situation is described in Figure \ref{Inssepa1} for the case $\gamma=1$ and Figure \ref{Inssepa2} for the case $\gamma=\frac{1}{2}$, respectively. In both figures,  the red circle is the separatrix and the black trajectory inside the separatrix is the critical trajectory. Optimal trajectories depicted in blue start from the point $(1,0)$ and end, loosing  optimality, on the critical trajectory.

The proof of Theorem \ref{Insideseparatrix} is presented in Appendix B. We give here the main ideas and discuss where the assumption $\gamma\geq \frac{1}{\sqrt{3}}$ is used. We consider the critical curve, i.e., with $\omega=\omega_c$ and $a=a_c$. Starting from the  point $(1,0)$ the distance from the origin decreases monotonically according to formula (\ref{rquadro}) and the last point is on the critical circle. Under the assumption $\gamma \geq \frac{1}{\sqrt{3}}$ the whole critical trajectory is in the first quadrant.\footnote{This renders the geometry of the problem easier to visualize. If this is not verified, the critical trajectory looks like a spiral winding around the origin more and more times as $\gamma \rightarrow 0$.} We introduce a parameter $\lambda:=\sin(a_c t)$, with $t \in [0, \frac{\pi}{2 a_c}]$, i.e., $\lambda \in [0,1]$, to parametrize the critical trajectory. Because of (\ref{rquadro}), $\lambda$ indicates the distance of the point on the critical trajectory from the origin, which goes from $1$
to $\sqrt{1-\frac{\gamma^2}{a_c^2}}$. For a given value of $\lambda$, i.e., for points on the same circle,  we compare the phase of any trajectory (corresponding to a given value of $\omega$ and $a$)
with the phase for the critical trajectory. In doing that, we assume $0 \leq t \leq \frac{\pi}{2a}$ and  we  use formula (\ref{fase1}) for the phase. We find that the phase for the generic trajectory is always bigger than the one for the critical trajectory, Lemma \ref{LemmaB1}. This has several consequences: 1) Every trajectory corresponding to $\omega \in (\omega^*,3\omega^*]$, $\omega \not= \omega_c$ that intersects the critical trajectory has to do so at a time $t > \frac{\pi}{2a}$ (Corollary \ref{CorollarioB1}). 2) All trajectories corresponding $\omega \in (2 \omega^*, 3 \omega^*]$ which under the assumption $\gamma \geq \frac{1}{\sqrt{3}}$ are also in the first and second  quadrant until $\frac{\pi}{2a}$, do not reach any of the  points below the critical curve and outside the critical circle (i.e no points in the region R of Figure \ref{Perlaprova}), before hitting the boundary of the unit disk at time $t:=\frac{\pi}{a}$ (Corollary \ref{CorollarioB2}).

We then consider a curve obtained from the critical trajectory by slightly modifying it lowering the phase by a small quantity $\epsilon \lambda$, for $\lambda\in [0,1]$. This curve is in the region below the critical curve and outside the critical circle. Since the trajectories with $\omega \in (2\omega^*, 3\omega^*]$ cannot reach them optimally (they touch the boundary of the unit disk first (Corollary \ref{CorollarioB2})), the only trajectories left are the ones in $[\omega^*,2 \omega^*)$. Because of the existence of the optimal control, there is at least one value $\omega \in  [\omega^*,2 \omega^*)$ and the corresponding trajectory which reaches the point corresponding to $\lambda$. Equating the two phases
up to a multiple of $2k\pi$, we find that, for a given $\lambda$, there exists a unique $\omega$ such that the two curve intersect at that $\lambda$ (Lemma \ref{Correspondence}). Here the assumption $\gamma \geq  \frac{1}{\sqrt{3}}$ is used to show that only the case $k=-1$ has to be used in the multiple $2k\pi$. In fact Lemma \ref{Correspondence} establishes a on to one  correspondence $\omega=\omega(\lambda)$ between the values $\lambda \in [0,1]$ and optimal values $\omega \in [\omega^*, 2\omega^*]$
As a consequence of the principle of optimality all points in $\mathcal{S}$ are reached by an optimal trajectory with $\omega \in [\omega^*, 2 \omega^*]$. This excludes however the points  between the critical curve and the $\epsilon$-deformed curve. However these points are recovered at the limit as $\epsilon \rightarrow 0$. This is in the conclusion of the proof of Theorem \ref{Insideseparatrix} in Appendix B.

\section{Discussion}\label{Geo}

The above analysis provides a description of the optimal trajectories for every element in $SU(2)$. It also gives a very simple method to find the optimal control for an element $X_f \in SU(2)$.

Given such element one first singles out the $(1,1)$ element and the point $P_f$ in the unit disk and checks whether $P_f$  is inside or outside the separatrix. If $P_f$ is outside one has to use the trajectories described in section \ref{operatoridiagonali}, i.e., with $\omega \in (-\infty, \omega^*)$. The choice of $\omega$ can be made by successive approximations (for example using a simple bisection algorithm)  by examining the plots for trajectories which leave $P_f$ on the right or on the left and getting closer and closer to the trajectory which actually contain $P_f$. If $P_f$ is inside the separatrix, the same procedure can be performed with the trajectories described in Section \ref{inside}. Once $\omega$ is found, one finds the corresponding $t$, either by tracing the plot or by solving an optimization problem minimizing (in $t$) the distance of the point on the trajectory from $P_f$. The last step is to adjust the phase $\tilde \phi$ in (\ref{soluzexpli}) (with the values found for $\omega$ and $t$) so that the element $(1,2)$ in (\ref{soluzexpli}) also coincides with the corresponding element in $X_f$. This completely determines the optimal controls in (\ref{controlli}).

\begin{figure}[htb]
\centering
\includegraphics[width=0.7\textwidth]{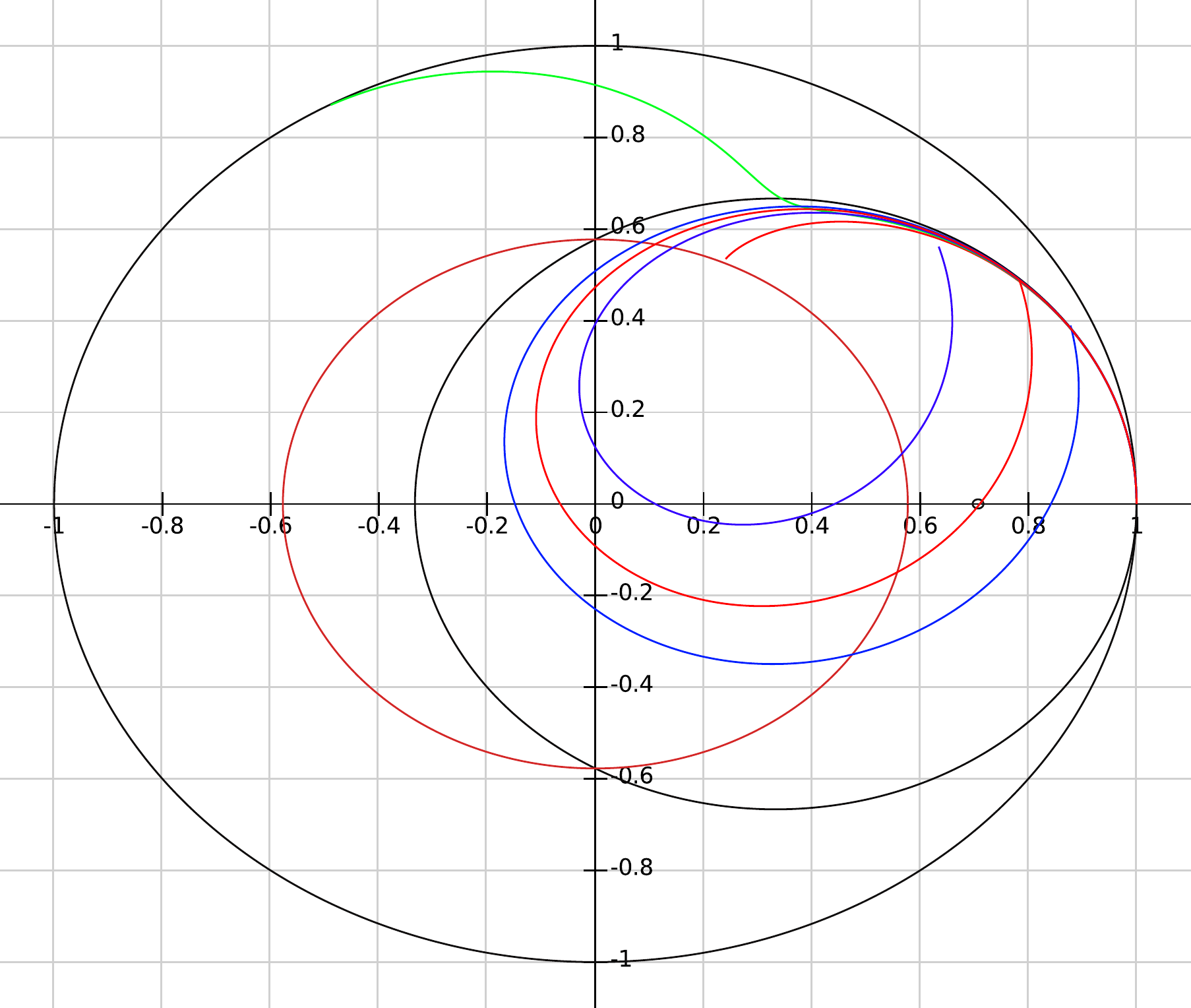}
\caption{Search for the optimal control for the Hadamard gate 
in the case $\gamma=\frac{1}{\sqrt{2}}$. The separatrix is in black and the critical trajectory and critical circle are in red. Two trial trajectories corresponding to $\omega=1.2*\omega^*$ and $\omega=1.4*\omega^*$ are in blue and the (approximate) optimal trajectory is in red. In green it is depicted a typical trajectory for $\omega>2\omega^*$. In this case it is $\omega=2.4\omega^*$. As predicted by the theoretical analysis, this trajectory follows the critical trajectory with higher phase getting close to the critical circle and then goes further away from the origin until reaching the boundary of the unit disk.}
\label{FigSUmmary}
\end{figure}

Figure \ref{FigSUmmary} describes the work we have done to find the optimal control for the Hadamard gate $X_f:=\frac{1}{\sqrt{2}}\left(\begin{array}{cc}
 1 & 1 \\ -1 & 1\end{array}\right)$ and $\gamma=\frac{1}{\sqrt{2}}$. The point $P_f$ is the point $(\frac{1}{\sqrt{2}}, 0)$ which is inside the separatrix curve. We have drawn a small circle around this point. The two curves in blue in the figure correspond to $\omega=1.2\omega^*$,  $\omega=1.4\omega^*$ and $\omega^*=\frac{3}{4}$ in this case. The optimal curve is found for $\omega \approx 1.28 \omega^*$ and is the curve in red crossing the small circle in the figure. The optimal time is found to be approximately $t_{opt}\approx \pi+0.2$. The total phase of the $(1,2)$ element in (\ref{soluzexpli}) must be zero, therefore, we choose $\tilde \phi=-\omega t_{opt}=1.28 \omega^*(\pi +0.2)$. These values have to be replaced in (\ref{controlli}) to give the optimal controls.\footnote{Recall that we have replaced the notation $\tau$ with $t$ therefore we should have $2t$ in (\ref{controlli}) instead of $t$.}

\vs


While we were completing this work, other authors \cite{newpaperbis} submitted a paper  on the same topic, building upon their previous work in \cite{newpaperonTimeOptimal} on the case $\omega_0=0$ (cf. Remark \ref{primobis}) and previous work in \cite{Boozer}, \cite{Kirillova}. In the paper \cite{newpaperbis}, the authors parametrize elements in $SU(2)$ with the so-called Hopf parameters\footnote{ These are defined as $\theta_{1,2,3}$ when writing $x=\cos(\theta_1) \cos(\theta_2)$, $y=\cos(\theta_1)\sin(\theta_2)$, $c=\sin(\theta_1)\cos(\theta_3)$, $d=\sin(\theta_1)\sin(\theta_3)$ (\ref{homeo}).} and the Euler parameters of the elements of $SU(2)$. They derive the dynamical equations in terms of these parameters and consider the optimal control problem in this setting. They prove properties of the optimal trajectories  and give an algorithm to find the optimal controls. Our geometric analysis of the optimal trajectories in the unit disk provides an alternative approach which, beside giving a very straightforward method to find the time optimal control, as we have seen above, highlights the general picture of the optimal trajectories. Main features of this picture are the existence of a closed curve which separates two classes of optimal trajectories (the separatrix) and of a special (non-smooth) trajectory inside this curve (the critical trajectory) which is some sort of limit of all other trajectories and it is where these trajectories loose optimality. It will be interesting in the future to investigate if, how and in what cases these features can be found in higher dimensional time optimal control systems on Lie groups.

\vs

\section*{Acknowledgement} D. D'Alessandro's research was supported by ARO MURI under Grant
W911NF-11-1-0268. D. D'Alessandro also acknowledges the kind hospitality of the Department
of Mathematics at the University of Padova, where part of this work was
performed. Graphs were drawn using  `fooplot' at
www.fooplot.com.

\section*{Appendix A}

\subsection*{Proof of Lemma \ref{Lemma1}}

\bpr
Using the expression of $a$ in terms of $\omega$, we have
\be{newTKM}
T_{k,m}= \frac{\pi k}{\sqrt{(1-\omega)^2+\gamma^2}},
\ee
where $\omega$ is chosen to satisfy the equation (cf. (\ref{equa2}))
subject to:
\be{1}
1+\frac{w}{ \sqrt{(1-\omega)^2+\gamma^2}}=\frac{\psi+2m\pi}{k\pi}.
\ee
Since the function $\frac{\omega}{a}$ has a maximum of $\frac{\sqrt{1+\gamma^2}}{\gamma}$ and an infimum at $-1$,   an $\omega \in \RR$ satisfying (\ref{1}) exists if and only if
\be{2}
0<\frac{\psi+2m\pi}{k\pi}\leq 1+\frac{\sqrt{1+\gamma^2}}{\gamma},
\ee
and there are at most two such $\omega$'s.

From now on we assume to have fixed a value for $k$ as in the statement of the Lemma. Set also $\alpha(m):=\frac{\psi+2m\pi}{k\pi}-1 \leq \frac{\sqrt{1+\gamma^2}}{\gamma}.$ $T_{k,m}$ is a decreasing function of $|b|:=|1-\omega|$.   Therefore the minimum is obtained for the value of $m$ which corresponds to the  maximum vale of $|b|$, satisfying
\be{4}
f(b):=\frac{1-b}{\sqrt{b^2+\gamma^2}}= \alpha(m).
\ee
The claim of the Lemma is proved if we show that this is obtained for $m=0$.
It is easy to see that:
\be{5}
-1< \alpha(0)=-1+\frac{\psi}{k\pi} <1,
\ee
and that
\be{6}
\alpha(m+1)=\alpha(m)+ \frac{2m}{k}.
\ee
The proof is based on the plot of the function $f(b)$ in Figure \ref{Figura2}, which is the case $\gamma=\frac{1}{2}$.


Using (\ref{5}), there exists (cf. Figure \ref{Figura2}) exactly one value  $b=b(0)$ such that $f(b(0))=\alpha(0)$, and since $f(0)=1\gamma \geq 1$ the value $b(0)$ is greater than zero.
Moreover as long as $\alpha(m)\leq 1$ we have that there exists only one value $b(m)$  such that $f(b(m))=\alpha(m)$, and we have $0 \leq b(m)<b(0)$.

The first $m$ such that $\alpha(m)>1$ is $m=k$, in which  case $\alpha(m):=\alpha(k)=1+  \frac{\psi}{k\pi}$. It is in fact easily seen that  $\alpha(k-1)=-1+\frac{\phi}{k\pi} +2-\frac{2}{k}=1+\frac{\phi-2\pi}{k} <1.$

From the plot  of the function $f$, it follows  that, for all $m>k$, if $b(m)$ is the solution with the maximum absolute value of $f(b(m))=\alpha(m)$, then $|b(m)|<|b(k)|$.

Thus to show that $T_{k,m}$ is minimum at $m=0$, we need only to show that
\[
|b(k)|<b(0),
\]
where $b(k)$ now denotes the solution of $f(b)=\alpha(k)$ of maximum absolute value which is assumed to be negative.\footnote{This follows from the plot of the function and if this was not the case, the claim would follow from the fact that the function $f(b)$ is decreasing for positive values of $b$.}
This  is equivalent to
\be{8}
-b(k)<b(0).
\ee
Since the function $f(b)$ is decreasing for $b>0$, if we show that $f(-b(k))>f(b(0))$, then equation (\ref{8}) follows. We compute:
\[
f(-b(k))=\frac{1+b(k)}{\sqrt{b(k)^2+\gamma^2}}= \frac{1-b(k)}{\sqrt{b(k)^2+\gamma^2}} +\frac{2b(k)}{\sqrt{b(k)^2+\gamma^2}}=
1+  \frac{\psi}{k\pi} + \frac{2b(k)}{\sqrt{b(k)^2+\gamma^2}}.
\]
Since $\left| \frac{b(k)}{\sqrt{b(k)^2+\gamma^2}}\right|<1$, we have that:
\[
f(-b(k))>1+  \frac{\psi}{k\pi} -2=f(b(0)),
\]
as desired.
\epr

\subsection*{Proof of Lemma \ref{Lemma2}}

\bpr
For a given $\psi \in (0, 2\pi)$, the expression of $T_{k,0}$ can be obtained from (\ref{equa1}), where $\omega$ and $a=a(\omega)$ are chosen so that (\ref{equa2}) is satisfied. In particular, after defining $x_{k,\psi}:=\frac{\psi - k \pi}{k \pi}$ and using the fact that $a$ is by definition positive, we obtain\footnote{Notice that $|x_{k,\psi}|<1$}
\be{aa}
a=\frac{-x_{k,\psi}+\sqrt{x_{k,\psi}^2+(1+\gamma^2)(1-x_{k,\psi}^2)}}{(1-x_{k,\psi}^2)}.
\ee
Replacing this and the expression of $x_{k,\psi}$ in $T_{k,0}=\frac{k \pi}{a}$, we obtain, after some algebra
\be{expressioTK}
T_{k,0}=T_{k,0}(\psi):=\frac{\psi(2 k\pi -\psi)}{k \pi - \psi + \sqrt{(k\pi)^2+\gamma^2 \psi (2 k \pi - \psi)}}.
\ee
We want to show that $T_{k,0}(\psi) > T_{1,0}(\psi)$ for every $\psi$. Since $\psi >0$ is a multiplicative
factor of very $T_{k,0}$ we can neglect it in comparing the two functions. Moreover since $\gamma$ is arbitrary, we can define $\gamma_1:=\gamma^2\psi>0$ and show, equivalently that $\tilde T_{k}(\psi)>\tilde T_{1}(\psi) $, for every $\psi$, with
\be{hhh}
\tilde T_{k}(\psi):=\frac{2 k \pi - \psi}{k \pi - \psi+ \sqrt(k\pi)^2+\gamma(2 k \pi - \psi)}.
\ee
Write $Y_k:=(k \pi)^2+ \gamma_1(2 k \pi - \psi)= Y_1+ \Delta_k$, with $\Delta_k:= (k^2-1) \pi^2+ 2 \pi \gamma_1(k-1)$, so that the claim is equivalent to
\be{u1}
(2k\pi -\psi)[(\pi - \psi)+ \sqrt{Y_1}] > (2 \pi - \psi)[k\pi - \psi + \sqrt{Y_1 + \Delta_k}].
\ee
After some algebra, we obtain
\be{u2}
(2 k \pi - \psi) \sqrt{Y_1} - (k-1) \pi \psi >  (2 \pi -\psi) \sqrt{Y_1 + \Delta_k}.
\ee
Since both sides are positive, we can square both sides, and collecting the terms containing $Y_1$, and using the definition of $\Delta_k$, we arrive at
\be{u3}
(4(k^2-1) \pi^2-4(k-1) \pi \psi) Y_1 +(k-1)^2 \pi^2 \psi^2-2(k-1)\pi \psi (2k \pi -\psi) \sqrt{Y_1}
>
\ee
$$(4 \pi^2 - 4\pi \psi + \psi^2)((k^2-1)\pi^2+ 2 \gamma_1 \pi (k-1)).$$
dividing everything by $\pi(k-1)$, we obtain,
\be{u4}
4 ((k+1) \pi - \psi) Y_1+(k-1)\pi \psi^2- 2 \psi (2k\pi - \psi) \sqrt{Y_1} > (4 \pi^2 - 4 \pi \psi + \psi^2)((k+1)\pi + 2 \gamma_1).
\ee
By collecting all terms that contain $Y_1$ on the left hand side and after some algebra, we obtain,
\be{u5}
(4(k+1)\pi - 4 \psi)Y_1-2 \psi (2 k \pi - \psi) \sqrt{Y_1} > 2 \pi \psi^2+4 \pi^2 (k+1)(\pi-\psi)+ 2 \gamma_1 (2 \pi -\psi)^2.
\ee
Using the expression for $Y_1$ (but not under the square root), we obtain, after some algebra,
\be{u6}
(2 k \pi - \psi) \gamma_1 (2 \pi - \psi)+ \pi \psi (2 k \pi -\psi)- \psi(2 k \pi - \psi) \sqrt{Y_1} >0,
\ee
which allows us to simplify the factor $(2 k \pi - \psi)$, so that the theorem is verified if
\be{u7}
\gamma_1(2 \pi - \psi) + \pi \psi > \psi \sqrt{Y_1}.
\ee
Taking the squares of both sides and reintroducing the expression of $Y_1$, we obtain, after some algebra,
\be{u8}
\gamma_1^2(2 \pi - \psi)^2 + 2 \pi \psi \gamma_1 (2 \pi - \psi) > \psi^2 \gamma_1( 2 \pi - \psi),
\ee
which after dividing by $\gamma_1 (2 \pi - \psi)$, gives
\be{u9}
\gamma_1 ( 2 \pi - \psi) + 2 \pi \psi > \psi^2.
\ee
This is certainly true for $\psi \in (0, 2 \pi)$ since $2\pi \psi > \psi^2$, which completes the proof.
\epr

\subsection*{Proof of Lemma \ref{stannofuori}}
\bpr
The optimal trajectories reaching the boundary of the unit disk do not intersect (before the boundary) because the intersection would mean that one of them is not optimal. Denote by $f(t,\omega):=(x(t, \omega), y(t,\omega))$ any  of these trajectories, parametrized by $\omega$, with $-\infty < \omega < \omega^*:=\frac{1+\gamma^2}{2}$.  The function $\omega$ as a function of $\psi$ (\ref{omegaopt}) is an increasing function of $\psi$. This implies that if $\omega_2 > \omega_1$  the curve $f(t,\omega_2)$ starts  below the curve $f(t,\omega_1)$, otherwise, they would have to  intersect.

Assume now by contradiction that the curve $f(t,\omega_1)$ at time $\bar t$ intersects the separatrix in the point $P:=f(t^*,\omega^*)$. Consider now a curve $f(t,\omega_2)$, with $\omega_2 > \omega_1$, and let $d_P$ denote the distance of $P$ from the curve  $f(t,\omega_2)$. Since there is no intersection between $f(t,\omega_1)$ and $f(t,\omega_2)$, $d_P>0$. Moreover for every $\omega > \omega_2$ the distance of the curve from $P$ is greater than $d_P$, otherwise there would be intersection of this curve with  $f(t,\omega_2)$. Consider now $f(t^*,\omega)$ and take the  limit $\lim_{\omega \rightarrow \omega^*} f(t^*,\omega)$, which by continuity must be $P$. However this contradicts the fact that the distance of any curve with $\omega > \omega_2$ from $P$ must be greater than $d_P>0$.

\epr
\section*{Appendix B: Proof of Theorem \ref{Insideseparatrix}}

Consider the critical trajectory,  for any $T\in [0,\frac{\pi}{2a_c}]$ let
\be{deflambda}
\lambda= \sin(a_cT),
\ee
then $\lambda \in[0,1]$. In the following we will use this variable $\lambda$ to parametrize the critical trajectory.

\bl{LemmaB1}
For any value of $\omega\neq \omega_c$, let $P(t)=(x(t),y(t))$ be a fixed point on the corresponding trajectory for $t\in (0,\frac{\pi}{2a}]$.
Then if there exists a $\lambda$ as in equation (\ref{deflambda}) such that the absolute value of $P(t)$ is equal to the absolute value of the point of the critical trajectory corresponding to $\lambda$, then
   the phase of $P(t)$ is strictly bigger than the phase of this point.
   \el
\bpr
Since the absolute values of the two points  are equal, we must have:
\[
\frac{\sin^2(at)}{a^2}=\frac{\lambda^2}{a^2_c},
\]
from this equation since $t\leq\frac{\pi}{2a}$, we derive:
\be{bfase1}
\sin(at)=\frac{a\lambda}{a_c}.
\ee
From which we have
\be{bfase2}
t=\frac{1}{a}\arcsin\left(\frac{a\lambda}{a_c}\right).
\ee
From equation (\ref{bfase1}), since $\cos(at)= \frac{\sqrt{a_c^2-a^2\lambda^2}}{a_c}$, we also have:
\be{bfase3}
\tan(at)= \frac{a\lambda}{\sqrt{a_c^2-a^2\lambda^2}}.
\ee
Let $\Phi_P(\lambda)$ be the phase of $P$, by using equation (\ref{fase1}) together with equations (\ref{bfase2}) and (\ref{bfase3}), we have that:
\be{bfase5}
\Phi_P(\lambda)= \omega\frac{1}{a}\arcsin\left(\frac{a\lambda}{a_c}\right) +\arctan\left( \frac{(1-\omega)\lambda}{\sqrt{a_c^2-a^2\lambda^2}}
\right).
\ee
Using the same argument, the phase $\Phi_c(\lambda)$ of the point of the critical trajectory corresponding to $\lambda$ is given by:
\be{fase6}
\Phi_c(\lambda)=  \omega_c\frac{1}{a_c}\arcsin(\lambda) + \arctan\left(\ \frac{\lambda(1-\omega_c)}{a_c \sqrt{1-\lambda^2}}
 \right).
\ee
We know that $\Phi_P(0)=\Phi_c(0)$, to prove that $\Phi_P(\lambda)>\Phi_c(\lambda)$ for $\lambda\in (0,1]$, we will prove that
 $\Phi'_P(\lambda)>\Phi'_c(\lambda)$.
 We have:
 \[
 \Phi'_P(\lambda)= \frac{\omega}{a}\frac{1}{\sqrt{\left(1-\frac{a^2\lambda^2}{a_c^2} \right)}} \frac{a}{a_c}+ \frac{a_c^2-a^2\lambda^2}{a_c^2-a^2\lambda^2+(1-\omega)^2\lambda^2}\frac{(1-\omega)(a_c^2-a^2\lambda^2)+(1-\omega)a^2\lambda^2}{(a_c^2-a^2\lambda^2)\sqrt{a_c^2-a^2\lambda^2}}=
 \]
\be{bfase7}
= \frac{\omega}{ \sqrt{a_c^2-a^2\lambda^2}} + \frac{(1-\omega)a_c^2}{ (a_c^2-\gamma^2\lambda^2)\sqrt{a_c^2-a^2\lambda^2}}
=\frac{1}{\sqrt{a_c^2-a^2\lambda^2}}\frac{a_c^2-\omega\gamma^2\lambda^2}{{a_c^2-a^2\lambda^2}}.
\ee
Moreover, we have:
\[
\Phi'_c(\lambda)=\frac{\omega_c}{a_c}\frac{1}{\sqrt{1-\lambda^2}} +\frac{a_c^2(1-\lambda^2)}{a_c^2(1-\lambda^2)+(1-\omega_c)^2\lambda^2}\frac{(1-\omega_c)}{a_c} \frac{(1-\lambda^2)\sqrt{1-\lambda^2}}{(1-\lambda^2)+\lambda^2}=
\]
\be{bfase8}
 =\frac{\omega_c}{a_c}\frac{1}{\sqrt{1-\lambda^2}} + \frac{(1-\omega_c)a_c}{\sqrt{1-\lambda^2} (a_c^2-\gamma^2\lambda^2)}
= \frac{a_c\sqrt{1-\lambda^2}}{ a_c^2-\gamma^2\lambda^2}.
 \ee
 Using equations (\ref{bfase7}) and (\ref{bfase8}), we have:
 \[
 \Phi'_P(\lambda)>\Phi'_c(\lambda)  \  \   \Leftrightarrow \   \
 \frac{a_c^2-\omega\gamma^2\lambda^2}{\sqrt{a_c^2-a^2\lambda^2}}> a_c\sqrt{1-\lambda^2}.
 \]
 Thus, we want to prove that:
 \be{bfase9}
a_c^2-\omega\gamma^2\lambda^2> a_c\sqrt{1-\lambda^2}\sqrt{a_c^2-a^2\lambda^2}.
\ee
By taking the squares we need to prove:
\[
a_c^4 +\omega^2\gamma^4\lambda^4-2\omega\gamma^2\lambda^2- a_c^2 (1-\lambda^2)(a_c^2-a^2\lambda^2)>0,
\]
which becomes:
\[
\lambda^4(\omega^2\gamma^4-a_c^2a^2)+\lambda^2(a_c^2a^2+a_c^4-2\omega\gamma^2a_c^2)=
\lambda^2\left( -\gamma^2(\omega-\omega_c)^2\lambda^2+a_c^2(\omega-\omega_c)^2\right)=
\]
\[
=\lambda^2(\omega-\omega_c)^2(a_c^2-\gamma^2\lambda^2)= \lambda^2(\omega-\omega_c)^2\gamma^2
(\gamma^2+1-\lambda^2)
>0,
\]
where the last equality holds since $\lambda<1$.
\epr

\bc{CorollarioB1}
Any trajectory corresponding to a value of $\omega$ and $a$, with $\omega\not=\omega_c$  cannot intersect the critical trajectory for $t \leq \frac{\pi}{2a}$.
\ec

\bpr
Assume, by contradiction, that there exists an $\omega\neq\omega_c$ and a time $t\in (0,\frac{\pi}{a}]$,
such that the  corresponding  trajectory intersect the critical one.   Denote by $P(t)=(x(t),y(t))$, the point of intersection.
Then there exists a $\lambda \in (0,1]$ such that the absolute value of $P(t)$ is
 equal to the absolute value of the point of the critical trajectory corresponding to $\lambda$.
 By applying Lemma \ref{LemmaB1},  we know that  the phase of $P(t)$ is strictly bigger than the phase of this point, so the two points
 are different.
\epr

\bc{CorollarioB2} Consider a trajectory corresponding to $\omega \in (2 \omega^*, 3 \omega^*]$. Such a trajectory never enters the critical circle. Moreover if it goes in the region below the critical trajectory, it does so at time $t > \frac{\pi}{a}$ (after it touches the boundary).
\ec

\bpr

Fix an $\omega \in (2 \omega^*, 3 \omega^*]$, and let $P(t)$  the point at time $t$
on the corresponding trajectory,  and $P_c(t)$ the one on the critical trajectory.

Since for $\omega>1$ $a=\sqrt{\gamma^2+(1-\omega)^2}$ is an increasing function of $\omega$, we know by
{\em Fact 1} (in Section \ref{prope}) that the absolute value of $P(t)$ is bigger than the absolute value of $P_c(t)$ for $t$ in a suitable neighborhood of $0$. By Lemma \ref{LemmaB1}, we also have that the phase of $P(t)$ is bigger than the one of $P_c(t)$.
Thus near $t=0$ the trajectory  corresponding to $\omega$  is  in the region above the critical curve.
Since, by  Corollary \ref{CorollarioB1}, we know that this trajectory   does not intersect the critical trajectory
for $t\in (0,\frac{\pi}{2a}]$, we have that in this  time interval it stays in the region above the critical curve.

For every $\omega$, we have
\[
|P(\frac{\pi}{2a})|^2=\frac{(1-\omega)^2}{a^2},
\]
thus  $|P(\frac{\pi}{2a})|>|P_c(\frac{\pi}{2a})|$.

Moreover, for $t\in [\frac{\pi}{2a}, \frac{\pi}{a}]$, the absolute value of $P(t)$ is increasing ({\em Fact 1},  Section \ref{prope}), thus
\be{bfase20}
 |P(t)|>|P(\frac{\pi}{2a})|>|P_c(\frac{\pi}{2a})|,
 \ee
  so the trajectory never enters the critical circle.

It remains to prove that also the trajectory for $t\in [\frac{\pi}{2a}, \frac{\pi}{a}]$ remains in the region above the critical curve, which is equivalent to saying  that it does not intersect the critical curve.

Consider $P(\frac{\pi}{2a})$ and let $\bar{t}\in (0,\frac{\pi}{2a_c})$ be such that $|P(\frac{\pi}{2a})|=|P_c(\bar{t})|$.
Let $\bar{\lambda} =\sin(a_c\bar{t})$ (see equation (\ref{deflambda})).

Since $|P(\frac{\pi}{2a})|=|P_c(\bar{t})|>|P_c(t)|$ for all $t\in[\bar{t},\frac{\pi}{2a_c}]$, we have that this second part of the trajectory does not intersect the critical one for $t>\bar{t}$.

The phase of $P(\frac{\pi}{2a})$ is bigger than the phase of $P_c(\bar{t})$, and so also of the phase of $P_c(t)$ for $t\in [0,\bar{t}]$.
 By {\em Fact 2},  in Section \ref{prope}, the phase
of $P(t)$ is bigger than the phase of $P(\frac{\pi}{2a})$, so:
\[
\texttt{  phase of }   P(t)  > \texttt{ phase  of  } P_c(t) \  t\in [0,\bar{t}].
\]
 Moreover, since $\gamma\geq1/\sqrt{3}$, it can be easily seen  that the critical curve lies in the first quadrant and
 that the phase of $P(t)$ is less that $\frac{3}{2}\pi$, so this second part of the trajectory does not intersect the critical one also for
  $0\leq t\leq \bar{t}$.

\epr

\vs

Now we slightly deform the critical trajectory so that, for every $\lambda$, the new trajectory, which is still parametrized by $\lambda\in [0,1]$,  is below the critical trajectory but still inside the separatrix and outside the critical circle. We call such a curve the {\it $\epsilon$-curve}.
The $\epsilon$-curve is obtained as follows. For every $\lambda$, the point on the curve has the same radius as the critical trajectory and phase $\psi_\epsilon(\lambda):=\psi(\lambda)-\epsilon \lambda$, where $\psi(\lambda)$ is the corresponding phase for the critical trajectory. Notice that   for $\lambda=0$, the phases are the same.

Given an $\epsilon$-curve, denote
     by $\zeta^{\epsilon}$ the
 map   which associates to every $\lambda \in (0,1)$ the unique $\omega_o^{\epsilon}(\lambda)$,
such that the point corresponding to $\lambda$ on the $\epsilon$-curve is reached by the trajectory corresponding to $\omega_o^{\epsilon}(\lambda)$  in minimum time.  Let also $\zeta^{\epsilon}(0)=\omega^*$.

The next Lemma proves that, for $\epsilon$ sufficiently small this map  is   a one-to-one, increasing map from $[0,1]$ to $[\omega^*, 2 \omega^*)$.

\bl{Correspondence}
Consider an $\epsilon$-curve with $\epsilon$ sufficiently small so that the curve is entirely contained in $\mathcal{S}$.
Let  $\zeta^{\epsilon}$ be the map defined above, then this map  is a   one to one, differentiable, and increasing, function from  $[0,1]$ to $ [\omega^*, \omega_c=2 \omega^*)$ .
 \el
\bpr
 Given an $\epsilon$-curve, fix a value ${\lambda}\in (0,1]$ and denote by ${P^{\epsilon}(\lambda)}$ the corresponding point on the
$\epsilon$-curve and by $P_c(\lambda)$ the the corresponding point on the
 critical curve. We have:
 \be{b640}
|P^{\epsilon}(\lambda)|=|P_c(\lambda)| \  \texttt{ and } \ \psi_{\epsilon}(\lambda)= \psi(\lambda)-\epsilon \lambda,
\ee
where  $ \psi_{\epsilon}(\lambda)$ denotes the phase of $P^{\epsilon}(\lambda)$ and
$ \psi(\lambda)$ denotes the phase of $P_c(\lambda)$.
 Since $ {P^{\epsilon}(\lambda)}$ is in the region below the critical curve and in $\mathcal{S}$, we know that $\omega^{\epsilon}_o(\lambda)\in (\omega^*,\omega_c)$.
 Since we have set $\zeta^{\epsilon}(0)=\omega^*$ the image of the function $\zeta^{\epsilon}$ is in the desired interval.

Next we   prove that this map $\zeta^{\epsilon}$ is  differentiable and strictly increasing, from which injectivity  follows.

Assume $\lambda\in(0,1]$,  denote by $a^{\epsilon}_0(\lambda)=\sqrt{\gamma^2+(1-\omega^{\epsilon}_0(\lambda))^2}$ and,
for $t\in[0,\frac{\pi}{a^{\epsilon}_0(\lambda)}]$,  by $P_o(t)$ the point at time $t$ in the trajectory corresponding to the control
$\omega^{\epsilon}_o(\lambda)$. 
Then, we must have:
\be{b642}
{P^{\epsilon}(\lambda)}=  P_o(t) \   \texttt{ for some } \in(\frac{\pi}{ 2a^{\epsilon}_0(\lambda)},\frac{\pi}{ a^{\epsilon}_0(\lambda)}],
\ee
 where we know that $t>\frac{\pi}{ 2a^{\epsilon}_0(\lambda)}$ by using Lemma \ref{lemmino}, Corollary \ref{CorollarioB1} and the fact that the phase of ${P^{\epsilon}(\lambda)}$ is less than the phase of $P_c(\lambda)$.

We have:
\[
|{P^{\epsilon}(\lambda)}|^2= | P_c(\lambda) |^2= 1-\frac{\lambda^2}{a_c^2},
\]
and
\[
| P_o(t) |^2= 1-\frac{\sin^2(a^{\epsilon}_o(\lambda)t)}{(a^{\epsilon}_o(\lambda))^2}.
\]
Equation (\ref{b642}), together with   the two previous  equations, since all quantities  are positive, implies:
\be{b643}
 \frac{\sin (a^{\epsilon}_o(\lambda)t)}{a ^{\epsilon}_o(\lambda)}=\frac{\lambda}{a_c}.
 \ee

Equation (\ref{b643}) gives a first relation between $\lambda$ and $\omega^{\epsilon}_o(\lambda)$. This relation involves also the variable $t$.
Next, by equating the phases of the two points ${P^{\epsilon}(\lambda)}$ and  $P_o(t)$ we will find another relation
between $\lambda$ and $\omega^{\epsilon}_o(\lambda)$, which will enable us to eliminate the $t$ dependence and find an implicit
formula of the type $F^{\epsilon}(\lambda, \omega^{\epsilon}_o(\lambda)=0$. From this relation and using the implicit map Theorem we will prove
our statement.

Using the definition of $\lambda$ given by equation (\ref{deflambda}),  the expression of the phase
given in   equation (\ref{fase1}), and the fact that the separatrix is in the first quadrant,  we have:
\be{b644}
\psi(\lambda)=\omega_c\frac{\arcsin(\lambda)}{a_c} +\arctan\left(\frac{(1-\omega_c)\lambda}{a_c\sqrt{1-\lambda^2}}\right).
\ee
Now, using equation (\ref{fase2}),  since $t>\frac{\pi}{ 2a^{\epsilon}_0(\lambda)}$, we also have:
\be{b645}
\texttt{  Phase } P_o(t) = \omega^{\epsilon}_o(\lambda)t +\pi +\arctan\left( \frac{(1-\omega^{\epsilon}_o(\lambda))}{(a^{\epsilon}_o(\lambda)}
\tan (a^{\epsilon}_o(\lambda)t)\right)
\ee
From equation (\ref{b643}),  and since $t\in [\frac{\pi}{ 2a^{\epsilon}_0(\lambda)},\frac{\pi}{ a^{\epsilon}_0(\lambda)}]$ we have:
\[
t=\frac{1}{a^{\epsilon}_o(\lambda)} \left(\pi-\arcsin\left(\frac{a^{\epsilon}_o(\lambda)\lambda}{a_c}\right)\right).
\]
Thus, using the previous equality and also equation (\ref{b643}),  we can rewrite equation (\ref{b645}) as:
\be{b646}
\texttt{  Phase } P_o(t) = \frac{\omega^{\epsilon}_o(\lambda)}{a^{\epsilon}_o(\lambda)} \left(\pi-\arcsin\left(\frac{a^{\epsilon}_o(\lambda)\lambda}{a_c}\right)\right) +\pi +\arctan\left(-  \frac{(1-\omega^{\epsilon}_o(\lambda))\lambda}{\sqrt{a_c^2-(a^{\epsilon}_o(\lambda))^2\lambda^2}}\right).
\ee
Since ${P^{\epsilon}(\lambda)}=  P_o(t)$, the  phases must be equal  up to a multiple of $2\pi$, thus we must have
\be{b647}
\psi_{\epsilon}(\lambda)  = \texttt{  Phase } P_o(t)+ 2k\pi,
\ee
for some $k\in\ZZ$. Since $P_c(\lambda)$ is in the first quadrant, we have
\[
-{\epsilon}\leq  \psi_{\epsilon}(\lambda)\leq \frac{\pi}{2}.
\]
Since $\gamma\geq \frac{1}{\sqrt{3}}$, we have that $1\leq \frac{\omega^{\epsilon}_o(\lambda)}{a^{\epsilon}_o(\lambda)} \leq 2$,
and since  the argument inside the function $\arctan$ in equation (\ref{b646}) is positive,  we have:
\[
\frac{3}{2}\pi \leq \texttt{  Phase } P_o(t) \leq \frac{7}{2}\pi.
\]
Given the previous bound for the two phases, the only possible $k\in \ZZ$ for which equality (\ref{b647}) holds is $k=-1$.
Thus we can rewrite equality (\ref{b647}), using $k=-1$ and equations (\ref{b644}) and (\ref{b646}), and we have:
\be{b6410}
\frac{\omega^{\epsilon}_o(\lambda)}{a^{\epsilon}_o(\lambda)} \left(\pi-\arcsin\left(\frac{a^{\epsilon}_o(\lambda)\lambda}{a_c}\right)\right) +\pi +\arctan\left(-  \frac{(1-\omega^{\epsilon}_o(\lambda))\lambda}{\sqrt{a_c^2-(a^{\epsilon}_o(\lambda))^2\lambda^2}}\right)-2\pi=
\ee
\[=
\omega_c\frac{\arcsin(\lambda)}{a_c} +\arctan\left(\frac{(1-\omega_c)\lambda}{a_c\sqrt{1-\lambda^2}}\right) -\epsilon\lambda.
\]
For $\omega\in [\omega^*,\omega_c)$ (and $a=\sqrt{(\gamma^2+(1-\omega)^2)}$)  and $\lambda\in [0,1]$, let:
\be{b649}
\begin{array}{cl}
F^{\epsilon}(\lambda, \omega)= &
 \frac{\omega}{a} \left(\pi-\arcsin\left(\frac{a\lambda}{a_c}\right)\right) - \arctan\left(   \frac{(1-\omega)\lambda}{\sqrt{a_c^2-a^2\lambda^2}}\right) -\pi - \\\
& \frac{\sqrt{(\gamma^2+1)}}{\gamma} \arcsin(\lambda) +\arctan\left(\frac{\gamma\lambda}{\sqrt{(\gamma^2+1)}\sqrt{(1-\lambda^2)}}\right)-\epsilon\lambda \end{array}
\ee
For any  $\lambda\in (0,1]$, and  corresponding $\omega_o^{\epsilon}(\lambda)$, equation (\ref{b6410}) says $F^{\epsilon}(\lambda, \omega_o^{\epsilon}(\lambda))=0$.  Moreover it  also holds that $F^{\epsilon}(0,\omega^*)=0$, so we have:
\be{b648}
F^{\epsilon}(\lambda,\zeta^{\epsilon}(\lambda))=0 ,  \  \ \texttt{ for all } \lambda\in [0,1].
\ee
We have:
\[
\frac{\partial F^{\epsilon}}{\partial \lambda}(\lambda,\omega)= \frac{-\omega}{\sqrt{(a_c^2-a^2\lambda^2)}}-\frac{(1-\omega)a_c^2}{\gamma^2
\left((\gamma^2+1)-\lambda^2\right)\sqrt{(a_c^2-a^2\lambda^2)}}+
\]
\[-  \frac{\sqrt{(\gamma^2+1)}}{\gamma}\frac{1}{\sqrt{1-\lambda^2}}
+\frac{\gamma\sqrt{(\gamma^2+1)}}{\sqrt{1-\lambda^2}\left((\gamma^2+1)-\lambda^2\right)}-\epsilon.
\]
Thus:
\be{b6411}
\frac{\partial F^{\epsilon}}{\partial \lambda}(\lambda,\omega)=
\frac{-(\gamma^2+1-\omega\lambda^2)}{
(\gamma^2+1-\lambda^2) \sqrt{(a_c^2-a^2\lambda^2)}}-
\frac{\sqrt{\gamma^2+1}\sqrt{1-\lambda^2}}{\gamma(\gamma^2+1-\lambda^2)} -\epsilon.
\ee
We also have:
\[
\frac{\partial F^{\epsilon}}{\partial\omega}(\lambda,\omega)=
\frac{\gamma^2+1-\omega}{a^3} \left(\pi-\arcsin\left(\frac{a\lambda}{a_c}\right)\right)+  \frac{\omega}{a^2}\frac{(1-\omega)\lambda}{\sqrt{a_c^2-a^2\lambda^2}}-\frac{\lambda}{\sqrt{a_c^2-a^2\lambda^2}}
\]
Thus:
\be{b6412}
\frac{\partial F^{\epsilon}}{\partial\omega}(\lambda,\omega)=
\frac{\gamma^2+1-\omega}{a^3} \left(\pi-\arcsin\left(\frac{a\lambda}{a_c}\right)\right)+\frac{\lambda(\gamma^2+1-\omega)}{a^2\sqrt{a_c^2-a^2\lambda^2}}
\ee
For any $\lambda \in (0,1]$ and $\omega \in [\omega^*, \omega_c)$, we have that:
\[
\frac{\partial F^{\epsilon}}{\partial\omega}(\lambda,\omega)>0
\]
Thus, we may apply   the Implicitly Mapping Theorem. Since our function
$\zeta^{\epsilon}$  describes the zeroes of the function $F^{\epsilon}$ (see equation (\ref{b648})) we can conclude that  for any $\lambda\in (0,1)$, $\zeta^{\epsilon}$ coincides locally with the implicit map. Thus,  $\zeta^{\epsilon}$ is differentiable, since $F^{\epsilon}$ is differentiable, and we have;
\[
(\zeta^{\epsilon})'(\lambda)=-\frac{\frac{\partial F^{\epsilon}}{\partial \lambda}(\lambda,\zeta^{\epsilon}(\lambda))}{\frac{\partial F^{\epsilon}}{\partial \omega}(\lambda,\zeta^{\epsilon}(\lambda))}>0,
\]
since, using equation (\ref{b6411}),  $\frac{\partial F^{\epsilon}}{\partial \lambda}(\lambda,\omega)<0$.

So for  $\lambda \in (0,1]$,  $\zeta^{\epsilon}$ is stricly increasing.
Now, to finish the proof, it is sufficient  to prove that
\[
\lim_{\lambda \to 0^+} \zeta^{\epsilon}(\lambda)=\omega^* =\zeta^{\epsilon}(0)
\]
By monotonicity we know that this  limit exists.

Let  $\lim_{\lambda \to 0^+} \zeta^{\epsilon}(\lambda)=\bar{\omega}$, then since $F^{\epsilon}(\lambda, \zeta^{\epsilon}(\lambda))=0$, and $F^{\epsilon}$ is continuous  we must have $F^{\epsilon}(0, \bar{\omega})=0$.
It holds:
\[
F^{\epsilon}(0,\bar{\omega})=\left(\frac{\bar{\omega}}{\bar{a}} -1\right) \pi,
\]
and this expression is zero if and only if $\frac{\bar{\omega}}{\bar{a}}=1$ which is equivalent to say $\bar{\omega}=\omega^*$.
So the Lemma is proved.
\epr

For any $\epsilon$, consider the map $\zeta^{\epsilon}$ and  define:
\be{b6413}
\omega^{\epsilon}_o(1):= \zeta^{\epsilon}(1)=\lim_{\lambda \to 1^-}\zeta^{\epsilon}(\lambda).
\ee
\bl{blemmanuovo}
We have:
\be{bnew1}
\lim_{\epsilon \to 0^+}{\omega^{\epsilon}_o(1)}=\omega_c.
\ee
\el
\bpr
We know that $\omega^*\leq \omega^{\epsilon}_o(1)\leq \omega_c$. Thus, without loss of generality we can assume that the limit for $\epsilon$ to $0^+$ of $\omega^{\epsilon}_o(1)$ exists, we denote this limit by $\omega^{0}_o(1)$.
Since $F^{\epsilon}(1, \omega^{\epsilon}_o(1))=0$, and it is continuos in the $\epsilon$ variable, we have:
\[
0= \lim_{\epsilon \to 0^+} F^{\epsilon}(1, \omega^{0}_o(1))= F^0(1, \omega^{0}_o(1))=
\]
\[=  \frac{\omega^{0}_o(1)}{a^0_o(1)} \left( \pi-\arcsin\left(\frac{a^0_o(1)}{a_c}\right) \right)-\arctan \left(\frac{1-\omega^{0}_o(1)}{\sqrt{a_c^2-a^0_o(1)^2}}\right) -\pi -\frac{\sqrt{\gamma^2+1}}{\gamma}\frac{\pi}{2}+\frac{\pi}{2}.
\]
It holds that $F^0(1, \omega_c)=0$, moreover since
\[
\frac {d}{dt} F^0(1,\omega) = \frac{\partial F^{\epsilon}}{\partial\omega}(1,\omega) >0,
\]
we have that $F^0(1,\omega)<0$ for $\omega^*\leq \omega<\omega_c$, so necessarily $\omega^{0}_o(1)=\omega_c$, as desired.
\epr

\subsection*{Conclusion of the proof of Theorem \ref{Insideseparatrix}}

From Lemma \ref{Correspondence} and Lemma \ref{blemmanuovo}, we have that there is a continuos,
one-to-one and onto correspondence between  controls in $[\omega^*,\omega_c]$ and points on the critical curve.
Define:
\be{concl10}
\begin{array}{cccc}
\zeta: & [0,1] & \to & [\omega^*,\omega_c] \\
 &      \lambda & \mapsto & \omega^0_o(\lambda),
 \end{array}
 \ee
with $ \omega^0_o(\lambda)= \lim_{\epsilon \to 0^+}{\omega^{\epsilon}_o(\lambda)}$. The function $\zeta$ is the uniform limit of the functions  $\zeta^{\epsilon}$.

Now we prove Theorem \ref{Insideseparatrix}. We need to prove:
\begin{enumerate}
\item
The trajectory corresponding to $\omega^*$ is the optimal for points of the separatrix.
\item
The trajectory corresponding to $\omega_c=2 \omega^*$ until the point (\ref{finalcritic}) is optimal for point on the critical trajectory.
\item
 For any other point in $\mathcal{S}$, there exists a unique value of $\omega \in (\omega^*, 2\omega^*)$ and an optimal trajectory corresponding to $\omega$ leading to that point.
\end{enumerate}

\vs

{\em{Proof of 1.}}

Using  Corollary (\ref{CorollarioB1}) we know that
any trajectory corresponding to a value of $\omega$ and $a$, with $\omega\not=\omega_c$   intersects  the critical trajectory after $ \frac{\pi}{2a}$.

 If $\omega\in (\omega^*,\omega_c)$, we have  $a_c>a$ (here the assumptions $\gamma\geq 1/ \sqrt{3}$ is used) thus the intersection is after $\frac{\pi}{2a_c}$.
 If $\omega=\omega^*$, then the separatrix does not intersect the critical curve.
 If $\omega>\omega^*$, all the trajectories loose their optimality after reaching the boundary and so before intersecting the critical curve.

{\em{Proof of 2.}}

Again if $\omega>\omega^*$, all the trajectories loose their optimality after reaching the boundary and so before intersecting the separatrix. Moreover, since the map $\zeta$ is onto, all the controls $\omega\in (\omega^*,\omega_c)$ intersect the separatrix after having intersected the critical curve and so these trajectory are no longer optimal.
Thus the separatrix must be optimal.

{\em{Proof of 3.}}

If the point we consider is below the critical curve or inside the critical circle, then the optimal control must be in $(\omega^*, 2\omega^*)$, so the statement holds.
Now we need also to prove that if the point is above the critical curve, the optimal control is still in $(\omega^*, 2\omega^*)$.
It holds that  all the trajectories corresponding to these controls are optimal, until they reach the critical curve.
The idea now is to prove that  any point
inside the separatrix, above the critical curve and outside the critical circle, is reached by one of them.
We do not write this part in details since the proof of this part follows the same line as the proof of statement 2. in Proposition \ref{Filling}.

\end{document}